\documentclass[%
 reprint,
 amsmath,amssymb,
 aps,
]{revtex4-1}

\usepackage{graphicx}
\usepackage{dcolumn}
\usepackage{bm}
\usepackage{multirow}         
\usepackage{color}
\usepackage{hyperref}         

\def\bea{\begin{eqnarray}}
\def\eea{\end{eqnarray}}
\def\be{\begin{equation}}
\def\ee{\end{equation}}
\def\fr{\frac}

\def\le{\left}
\def\ri{\right}

\begin{document}

\title{Bound Dark Energy: towards understanding the nature of the Dark Energy}
\author{Erick Almaraz}
 \email{ealmaraz@estudiantes.fisica.unam.mx, erickalmaraz@gmail.com}
\author{Axel de la Macorra}%
 \email{macorra@fisica.unam.mx, a.macorra@gmail.com}
\affiliation{%
 Instituto de F\'isica, Universidad Nacional Aut\'onoma de M\'exico, Ciudad de M\'exico, 04510 Mexico
}%

\date{\today}

\begin{abstract}
We present a complete analysis of the observational constraints and cosmological implications of our Bound Dark Energy (BDE) model aimed to explain the late-time cosmic acceleration of the universe. BDE is derived from particle physics and corresponds to the lightest meson field $\phi$ dynamically formed at low energies due to the strong gauge coupling constant. The evolution of the dark energy is determined by the scalar potential $V(\phi)=\Lambda_c^{4+2/3}\phi^{-2/3}$ arising from non-perturbative effects at a condensation scale $\Lambda_c$ and scale factor $a_c$, related each other by $a_c\Lambda_c/\mathrm{eV}=1.0934\times 10^{-4}$. We present the full background and perturbation evolution at a linear level. Using current observational data, we obtain the constraints $a_c=(2.48 \pm 0.02)\times10^{-6}$ and $\Lambda_c=(44.09 \pm 0.28) \textrm{ eV}$, which is in complete agreement with our theoretical prediction $\Lambda_c^{th}=34^{+16}_{-11}\textrm{ eV}$. The  BDE equation of state  $w_\mathrm{BDE} =p_\mathrm{BDE} /\rho_\mathrm{BDE}$ is a growing function at late times with  $-1<w_\mathrm{BDE}(z) < -0.999$ $(-0.950)$ for $132> z \geq 1.8$ $(0.35)$. The bounds on the equation of state today, the dark energy density and the expansion rate are $w_\mathrm{BDE 0}=-0.929\pm 0.007$, $\Omega_\mathrm{BDE0}=0.696\pm0.007$ and $H_0=67.82\pm 0.05$ km s$^{-1}$Mpc, respectively.  Even though the constraints on the six Planck base parameters are consistent at the 1$\sigma$ level between BDE and the concordance $\Lambda$CDM  model, BDE  improves the likelihood ratio by 2.1 of the Baryon Acoustic Oscillations (BAO) measurements with respect to $\Lambda$CDM and has an equivalent fit for type Ia supernovae and the Cosmic Microwave Background data. We present the constraints on the different cosmological parameters, and particularly we show the tension between BDE and $\Lambda$CDM in the BAO distance ratio $r_\mathrm{BAO}$ vs $H_\mathrm{0}$ and the growth index $\gamma$ at different redshifts, as well as the dark matter density at present time $\Omega_ch^2$ vs $H_0$. These results allow us to discriminate between these two models  with  more precise cosmological observations including distance measurements and large scale structure data in the near future.
\end{abstract}

\maketitle

\section{Introduction.}\label{introduction}
The observational evidence gathered during the last two decades \cite{Riess98_SNeIa,*Perlmutter98_SNeIa,Kowalski08_SNeIa,*Amanullah10_SNeIa,Betoule14_SNeIa,Eisenstein05_BAO,*Cole05_BAO,Blake11_BAO,Beutler11_BAO,Ross15_BAO,GilMarin16_BAO,Fosalba03_CMB,*Fosalba04_CMB,*Bough04_CMB,*Nolta04_CMB,*PlanckISW15_CMB,*deBernardis00_CMB,*WMAP9yr13_CMB,Tegmark06_LSS} shows that the expansion rate of the universe is accelerating. Within the framework of the standard cosmological model ($\Lambda$CDM), the dominant contribution to the energy content of the universe at present comes from the dark energy, which is described by a cosmological constant ($\Lambda$) in space and time and is responsible for the cosmic acceleration \cite{Gron18_CC}. The present cosmic budget is distributed into nearly 70\% of $\Lambda$ and 26\% of Cold Dark Matter (CDM), while the remaining 4\% consists of the Standard Model (SM) particles, mainly baryons, photons and neutrinos \cite{PlanckCP15_CMB}. Although the $\Lambda$CDM model is so far the simplest theoretical scheme describing consistently the evolution of the universe at large scales, the nature of the dark energy is still a mystery \cite{Weinberg89_CC,Martin12_CC,Sola13_CC,Gron18_CC}. The theoretical estimations of $\Lambda$  need to be unnaturally tuned up to 30 orders of magnitude \cite{Martin12_CC} to reproduce the observed value of $\Lambda$, which also leads to similar amounts of dark energy and dark matter at present. These two open issues of the standard $\Lambda$CDM model, commonly referred to as the fine-tuning and the coincidence problems, strongly suggest to go beyond $\Lambda$ and look for alternative models to elucidate the nature of the dark energy.

There has been gathered a great deal of observational evidence \cite{Weinberg13_DE} aimed to determine the properties and the dynamics of the dark energy, since the cosmic acceleration was firstly observed in measurements of the luminosity distance of type Ia supernovae (SNeIa) twenty years ago \cite{Riess98_SNeIa,*Perlmutter98_SNeIa}. Now we can look for the imprints of the dark energy on the Baryon Acoustic Oscillations (BAO) \cite{Bassett09_BAO} and the Large Scale Structure (LSS) of the universe \cite{Tegmark06_LSS}, the Cosmic Microwave Background radiation (CMB) \cite{Nishizawa14_CMB,PlanckDE15_CMB}, the lensing of the light coming from distant galaxies \cite{Kilbinger15_WL,*Mandelbaum18_WL} and the local determination of the current expansion rate of the universe ($H_0$) \cite{Riess16_H0}, among other probes. The information provided by all these measurements allows us to set tight constraints on the dark energy properties, explore the parameter space of alternative models and discriminate between different cosmological scenarios \cite{PlanckCP15_CMB,PlanckDE15_CMB}. Among these alternative scenarios we include: quintessence models \cite{Copeland06_QCDM,*Tsujikawa13_QCDM}, modifications of general relativity \cite{Clifton12_MG,*Bamba12_DE}, Chaplygin gases and other dark energy-dark matter interaction schemes \cite{Li18_Chaplygin,*Salvatelli14_DE,*Ferreira17_DE}, running vacuum \cite{Sola17_DE} and backreation models \cite{Rasanen04_DE,*Kolb11_DE}. In quintessence models \cite{Copeland06_QCDM,*Tsujikawa13_QCDM} the general approach is to replace the cosmological constant term by a scalar field $\phi$ whose dynamical evolution is determined by the potential $V(\phi)$. Unlike the $\Lambda$CDM scenario where the dark energy has a constant homogeneous energy density, in quintessence models we must take into account the dark energy perturbations, which leave imprints on the evolution of matter and radiation inhomogeneities leading to potentially detectable signals in the power spectrum \cite{Brax00_QCDM,Weller03_QCDM,*Bean04_QCDM}. Moreover, it is possible to find potentials such as $V(\phi)=M^{4+n}\phi^{-n}$ where the late-time evolution of the scalar field does not strongly depends on the initial conditions \cite{Steinhardt99_QCDM}, thus ameliorating the coincidence problem of $\Lambda$CDM. Early works on scalar fields as dark energy candidates were done by \cite{Peebles88_QCDM,*Ratra88_QCDM,*Wetterich88_QCDM} and subsequently by \cite{Caldwell98_QCDM,Steinhardt99_QCDM,Ma99_QCDM,Brax00_QCDM,Piccinelli00_QCDM,Weller03_QCDM,*Bean04_QCDM}. However, until recent times it has been possible to find accurate constraints and perform large complex non-linear simulations for these models \cite{Yashar09_QCDM,*Alimi10_QCDM,*Wang12_QCDM,*Gupta12_QCDM,*Chiba13_QCDM,*Takeuchi14_QCDM,*Chen15_QCDM,*Lima16_QCDM,*Smer17_QCDM,*Sola17_QCDM,*Bag18_QCDM}.\\\indent
In this paper we study in detail the Bound Dark Energy model presented in \cite{Macorra18_BDE}, where the dark energy corresponds to a light scalar meson particle dynamically formed at late times. Our model introduces a supersymmetric $SU(N_c=3)$ Dark Gauge group (DG) with $N_f=6$ massless particles. We further assume that the gauge coupling constant of the DG is unified with the couplings of the (supersymmetric) Standard Model (MSSM) at the unification scale $\Lambda_{gut}$. This assumption allow us tor reduce the number of free parameters of our model. For energies below  $\Lambda_{gut}$ the particles of the DG and the MSSM interact only through gravity. As the universe expands and the temperature drops off, the gauge coupling of the DG becomes strong and the DG particles form composite states whose mass is proportional to the condensation scale $\Lambda_c$, as similarly occurs with the pions, the protons and neutrons in the SM. The Dark Energy is the lightest scalar meson formed by the non-perturbative dynamics of the DG interaction, and its further evolution is described by a canonical scalar field $\phi$ with an inverse power law potential (IPL) $V(\phi)=\Lambda_c^{4+2/3}\phi^{-2/3}$ derived from the Affleck-Dine-Seiberg (ADS) superpotential \cite{ADS85_HEP}. Since in our model the dark energy arises from the binding of free particles, we refer to it as Bound Dark Energy (BDE).

This is a completely different scenario from other quintessence theories \cite{Copeland06_QCDM,*Tsujikawa13_QCDM} where the scalar field describing the dark energy is a fundamental field in nature. The full evolution of the quintessence field depends not only on the scalar potential $V(\phi)$, but also on the initial conditions of $\phi$ and $\partial \phi/\partial x^{\mu}$. For example, in IPL models $V(\phi)=M^{4+n}\phi^{-n}$ the parameters $M$ and $n$ of the scalar potential as well as the initial conditions of $\phi$ and $\partial\phi/\partial x^\mu$ are free parameters which have to be tuned to give the correct cosmological observations. On the other hand, in our BDE model the exponent $n=2/3$ of the scalar potential $V(\phi)=\Lambda_c^{4+2/3}\phi^{-2/3}$ is determined by the number of colors ($N_c=3$) and flavors ($N_f=6$) of the DG, and assuming gauge coupling unification at $\Lambda_{gut}$, the condensation scale $\Lambda_c$ and the scale factor $a_c$ setting the onset of BDE are all derived quantities \cite{Macorra03_BDE,Macorra18_BDE}. Moreover, since the initial conditions of the scalar field at the onset of BDE naturally arise from physical considerations, the content of dark energy at any time is simply determined by the solution of the background evolution equations. This means that our BDE model predicts the amount of dark energy today and therefore we have one less free parameter than $\Lambda$CDM.

IPL potentials derived from the ADS techniques were firstly obtained by \cite{Binetruy99_HEP}. However, the implications on the dynamics of the dark energy and the phenomenological acceptable scenarios were thoroughly studied some years later \cite{Macorra01_BDE,Macorra02_BDE,Macorra03_BDE,Macorra05_BDE}. In the following, we present the first precision constraints on the BDE model and how they depart from the $\Lambda$CDM scenario. We find that BDE agrees well with current observational data and the constraints on the six Planck base parameters are consistent at the $1\sigma$ level between BDE and standard $\Lambda$CDM. However, the different amount of dark energy and the presence of dark energy inhomogeneities in BDE leads to discrepancies w.r.t $\Lambda$CDM. Interestingly, BDE improves the fit to BAO measurements specially intended to determine the dynamics of the dark energy by increasing the likelihood ratio by 2.1 w.r.t. $\Lambda$CDM, and it has an equivalent fit for SNeIa and CMB data. This is reflected in tensions between BDE and $\Lambda$CDM in the BAO distance ratio $r_\mathrm{BAO}$ vs $H_0$ and the growth index $\gamma$ at different redshifts.

This paper is organized as follows. In section \ref{bde} we present the foundations of the BDE model. We discuss the cosmological evolution of the dark group before the phase transition and we derive some important results relating the relevant parameters of our model. We review the derivation of the IPL potential and then we present the equations of motion of the scalar field for the homogeneous part and its perturbations in linear theory. We show the generic solution illustrating the dynamics of the dark energy in our model. In section \ref{mcmc} we present the constraints on the model obtained from SNeIa, BAO and CMB measurements. The cosmological predictions based on these constraints are analyzed in section \ref{implications}, where we discuss the potential departures from the $\Lambda$CDM scenario that may be detectable in the near future. Finally, we summarise our findings and state our conclusions in section \ref{conclusions}. In this paper we adopt the usual conventions on notation and terminology found in the literature: a zero subfix (superfix) denotes the value of a given quantity at the present epoch and $a$ is the scale factor related to the cosmological redshift $z$ by $1+z=a_0/a$, where $a_0=1$ for a flat Friedmann-Lema\^itre-Robertson-Walker (FLRW) metric.

\section{The Bound Dark Energy model.}\label{bde}
\subsection{General  Framework}
The standard model of cosmology $\Lambda$CDM assumes Cold Dark Matter and a cosmological constant $\Lambda$ as the source for the current acceleration of the universe. The  cosmological constant $\Lambda$ has an equation of state $w_\Lambda \equiv -1$ and its energy density $\rho_\Lambda \equiv \Lambda/(8\pi G) $ is constant in time and space.  $G$ is the gravitational constant related to the Planck mass $m_\textrm{Pl}\equiv 1/\sqrt{G}$ in natural units. However, there is up to date no explanation on the origin nor on the magnitude of $\Lambda$ and it must be fine tuned by observations to an incredible one part in $10^{30}$, since the ratio of the observed value of $\rho_\Lambda$ and $m_\mathrm{Pl}$ is $\rho_\Lambda^{obs}/m_\mathrm{Pl}^4\simeq 10^{-120}$ \cite{Weinberg89_CC,Martin12_CC}. Alternative to $\Lambda$, scalar fields (quintessence) have been proposed to parametrize the Dark Energy (DE) and in the last decade a large number of quintessence models have been studied \cite{Yashar09_QCDM,*Alimi10_QCDM,*Wang12_QCDM,*Gupta12_QCDM,*Chiba13_QCDM,*Takeuchi14_QCDM,*Chen15_QCDM,*Lima16_QCDM,*Smer17_QCDM,*Sola17_QCDM,*Bag18_QCDM}. In the context of particle physics there are two different ways to generate these scalar fields: they  can be either fundamental particles ---such as the Higgs field---, or they can be composite particles made out of the fundamental quarks. The mass of a fundamental particle is a free parameter, whereas the mass of a composite state can be in principle related to the symmetry breaking scale, e.g., the mass of the pions is $m_\pi=140$ MeV, while the QCD scale is $\Lambda_\textrm{QCD}=210$ MeV \cite{PDG18_HEP}.

Here will follow the second case in which the DE is a composite scalar field generated at late times due to a strong gauge coupling constant of a hidden gauge group, and elaborate in the cosmological properties of the dark energy model present in \cite{Macorra18_BDE} referred to as Bound Dark Energy ---since the mass of the particle is due to the binding energy of the underlying gauge theory.  The BDE model introduces a supersymmetric dark gauge group $SU(N_c)$ with $N_c=3$ colors and $N_f=6$ massless flavors in the fundamental representation  \cite{Macorra01_BDE,Macorra02_BDE,Macorra03_BDE,Macorra05_BDE,Macorra18_BDE}. The fundamental status of $N_c$ and $N_f$ is the same that the fundamental status of the input parameters of the SM ---$SU_\textrm{QCD}(3)\times SU(2)_L\times U_Y(1)$ with 3 families describing the strong and electroweak interactions--- in the sense that they are quantities not derived from a deeper theory. There is up  to date no understanding from first principles on the choice of gauge groups  neither in the SM nor in BDE. Moreover, since $N_c$ and $N_f$ take integer values they cannot be fine tuned and are therefore they are  not cosmological  parameters.  Furthermore, we assume that the DG and the SM gauge groups are unified at the unification scale $\Lambda_{gut}\approx  10^{16}$ GeV as motivated by grand unification models, and below this scale the particles of the DG and the SM interact only through gravity \cite{Macorra03_BDE}. 
 
At high energies the DG particles are massless and weakly coupled, so they contribute to the total amount of radiation of the universe and its energy density redshifts as $\rho_\mathrm{DG}\propto a^{-4}$. However, the strength of the gauge coupling constant $g^2$ evolves with the energy and we can use the renormalization group equation to estimate its evolution. The coupling constant increases at lower energies and eventually it becomes strong at the condensation energy scale $\Lambda_c$ and at a scale factor denoted by $a_c$. At this scale gauge invariant states are created forming gauge neutral particles, i.e. dark mesons and dark baryons---similar as in the SM QCD force, where mesons (e.g. pions) and baryons (e.g. protons and neutrons) are formed at $\Lambda_\mathrm{QCD}=210\pm 14$ MeV. Strong gauge interactions produce a non-perturbative scalar potential $V(\phi)$ below $\Lambda_c$. The scalar potential $V(\phi)$ can be computed from the ADS techniques \cite{ADS85_HEP} and it is a function of the effective scalar field $\phi$, i.e., our BDE particle. Finally, the dynamical evolution of $\phi$ is determined by the scalar potential $V(\phi)$ and the initial conditions of $\phi$.

In particle physics there are two dynamically different ways to generate particle masses, namely, the Higgs mechanism and a non-perturbative gauge mechanism. In the SM the elementary particles (quarks, electrons, neutrinos) get their mass by the interaction with the Higgs field. The dynamically evolution of the Higgs field implies that at high energies all SM masses vanish, but once the Higgs settles into the minimum of its potential at the electroweak scale $\Lambda_\mathrm{EW} = \mathcal{O}(100\textrm{ GeV})$, the Higgs field acquires a non vanishing vacuum value giving a mass to the SM particles. Therefore, the mass of the fundamental particles vanishes at high energies and are non zero below the phase transition scale $\Lambda_\mathrm{EW}$. On the other hand, the non-perturbative gauge mechanism is based on the strength of gauge interaction, which evolves as a function of the energy. In the model we present here, the strength of the gauge coupling increases with decreasing energy and it becomes strong at the condensation scale $\Lambda_c$ at a scale factor $a_c$. The mass of BDE is proportional to $\Lambda_c$ at $a_c$, but it decreases as the universe expands as $m_\phi \propto \le( \Lambda_c/\phi \ri)^ {4/3} \Lambda_c$.

\subsection{Evolution of the gauge coupling, relativistic regime and condensation scale.}
The strength of the gauge coupling constant $g(E)$ evolves with the energy as determined by the renormalization group equation, and at one loop it is simply given by:
\begin{equation}\label{eq:bde_coupling}
g^{-2}(E) = g^{-2}(E_i)+ \frac{b_0}{8\pi^2} \ln \le(\fr{E}{E_i}\ri), 
\end{equation}
where $g_i \equiv g(E_i)$ corresponds to the value of the gauge coupling at some scale $E_i$ and $b_0$ counts the number of elementary particles charged under the gauge group. If $b_0 > 0$, as for the QCD strong force and our DG, the gauge coupling constant increases with decreasing energy $E$ and we have a non-abelian asymptotic free gauge group. The condensation scale or phase transition scale is defined as the energy when the coupling constant becomes strong, i.e. $g(E)\gg 1$, and from Eq. (\ref{eq:bde_coupling}) we have:
\begin{equation}\label{eq:bde_LLc}
\Lambda_c \equiv E_c = E_i e^{-8\pi^2/(b_0 g_i^2)}. 
\end{equation}
The fact that $\Lambda_c$  is exponentially suppressed compared to $E_i$ allows us to understand why $\Lambda_c$ can be much smaller
than the initial $E_i$, which may be taken as the Planck or the Unification scale $\Lambda_{gut}$. In our BDE model we have $b_0=3N_c-N_f=3$, with $N_c=3,N_f=6$. 

The condensation scale $\Lambda_c$ sets the phase transition scale, where above $\Lambda_c$ the elementary particles of the DG are massless and below $\Lambda_c$ the strong force binds these elementary fields together forming neutral bounds states as occurs with the  mesons and baryons in QCD. At high energies, the massless elementary fields of the DG are weakly coupled and they contribute to the total amount of radiation of the universe by: 
\begin{equation}\label{eq:bde_rhogT1}
\rho_\mathrm{DG}(a)=\frac{\pi^2}{30}g_\mathrm{DG}T_\mathrm{DG}^4(a),
\end{equation}
where $T_\mathrm{DG}$ is the temperature and $g_\mathrm{DG}$ are the relativistic degrees of freedom. It is common to express these extra relativistic degrees of freedom in terms of the extra number of effective neutrinos $N_{ext}$ as $\rho_{ext}=(\pi^2/30)(7/4) N_{ext} T_\nu^4$ \cite{PlanckCP15_CMB}, where $T_\nu$ is the neutrino temperature. Comparing this expression with Eq. (\ref{eq:bde_rhogT1}) we have:
\begin{equation}\label{eq:bde_rdg_1}
\rho_\mathrm{DG}=\fr{\pi^2}{30}\,\le(\fr{7}{4}\ri)\, N_{ext} \,\,T_\nu^4, 
\end{equation}
\begin{equation}\label{eq:bde_rdg_2}
N_{ext} \equiv  \le(\fr{4}{7}\ri) \,g_\mathrm{DG}\, \le(\fr{T_\mathrm{DG}}{T_\nu}\ri)^4.
\end{equation}

\subsection{Initial Conditions  and Gauge Coupling Unification}
The energy scale $\Lambda_c$ where the phase transition takes place depends on the values of the gauge coupling constant $ g_i=g(E_i)$
at $E_i$ and on $b_0=3N_c-N_f=3$. Motivated by grand unification theories we propose to unify our DG with the gauge groups of the minimal supersymmetric SM. The unification of the coupling constants in the MSSM model takes place at the unification scale $\Lambda_{gut}=(1.05\pm0.07)\times 10^{16}\textrm{ GeV}$ with $g_{gut}^2=4\pi/(25.83\pm 0.16)$ the unified coupling constant \cite{Bourilkov15_HEP}. Using these values of $\Lambda_{gut}$ and $ g^2_{gut}$ in the one loop renormalization group equation (\ref{eq:bde_LLc}) for our DG, we obtain a condensation scale \cite{Macorra03_BDE,Macorra05_BDE}:
\begin{equation}\label{eq:bde_Lc_theory}
 \Lambda_c=\Lambda_{gut} e^{-8\pi^2/(b_0 g _{gut}^2)} = 34^{+16}_{-11}\textrm{ eV}.
\end{equation}
The condensation scale  $\Lambda_c$ is no longer a free parameter in our model. Notice that the errors in  $\Lambda_c$ in Eq. (\ref{eq:bde_Lc_theory}) come from to the poorly constrained values of $\Lambda_{gut}$ and $g_{gut}^2$, mainly  due to  the uncertainties of the QCD gauge coupling constant \cite{Bourilkov15_HEP}.

At high energies all the particles of the DG and the MSSM are relativistic and the energy density can be expressed in terms of the relativistic degrees of freedom ($g$) and the temperature ($T$) as in Eq. (\ref{eq:bde_rhogT1}). For the MSSM we have:
\begin{equation}\label{eq:bde_rhogT_2}
\rho_\mathrm{SM}(a)=\frac{\pi^2}{30} g_\mathrm{SM} T_\mathrm{SM}^4(a).
\end{equation}

Gauge coupling unification implies that $T_\mathrm{DG}(a_{gut})=T_\mathrm{SM}(a_{gut})$ at $\Lambda_{gut}$, leading to the ratio:
\begin{equation}\label{eq:bde_rhoDGrhoSM_gut}
\frac{\rho_\mathrm{DG}(a_{gut})}{\rho_\mathrm{SM}(a_{gut})}=\frac{g_\mathrm{DG}^{gut}}{g_\mathrm{SM}^{gut}}=0.426,
\end{equation}
where $g_\mathrm{SM}^{gut}=228.75$ for the MSSM \cite{Macorra03_BDE} and: 
\begin{equation}\label{eq:bde_gDG}
g_\mathrm{DG}^{gut}=\left(1+\frac{7}{8}\right)\left[ 2(N_c^2-1)+2N_cN_f\right]=97.5,
\end{equation}
for the DG \cite{Macorra03_BDE}. This means that the DG accounts for the $\Omega_\textrm{DG}(a_{gut})=\rho_\textrm{DG}/(\rho_\textrm{SM}+\rho_\textrm{DG})=0.299$ of the cosmic budget of the universe at that time. 

Below the unification scale, the DG particles  interacts with the SM particles only through gravity and therefore the thermal equilibrium between these two groups is no longer maintained. Since the DG particles remain massless above energies $\Lambda_c$, the number of relativistic degrees of freedom $g_\mathrm{DG}$ is constant until the condensation epoch at $a_c$,  while the relativistic degrees of freedom $g_\mathrm{SM}$ of the SM decrease as the universe expands and cools down. We can use entropy conservation $S_x=\tfrac{2\pi^2}{45}g_xT_x^3a^3$ for $x=\mathrm{DG, SM}$ to relate their temperatures at different times. We find convenient to use the temperature of the neutrinos $T_\nu$ as reference for $T_\mathrm{SM}$, since they are in thermal equilibrium with photons for $T > 1 \textrm{ MeV}$  and after neutrino decoupling electrons and positrons annihilate, heating up the photon ($\gamma$) bath slightly above neutrino temperature, $T_\gamma = \left( \tfrac{11}{4} \right)^{1/3}T_\nu$. For $a\leq a_c$ entropy conservation leads to: 
\begin{equation}\label{eq:bde_TDGTSM_grl}
\frac{T_\mathrm{DG}(a)}{T_\mathrm{\nu}(a)}=\left( \frac{g_\mathrm{SM} (a)} {g_\mathrm{SM}^{gut} } \right)^{1/3}.
\end{equation}
We can use  Eq. (\ref{eq:bde_TDGTSM_grl}) to determine the temperature ratio $T_\mathrm{DG}/T_\mathrm{\nu}(a)$ at different values of $a<a_c$. For example, at neutrino decoupling at $a_{\nu dec}$ we find the ratio $T_\mathrm{DG}(a_{\nu dec})/T_\mathrm{SM}(a_{\nu dec})=g_\mathrm{SM}^{\nu dec}/ {g_\mathrm{SM}^{gut} }=  0.047$, where $g_\mathrm{SM}^{\nu dec}=2+\frac{7}{8}[2(3)+2(2)]=10.75$ accounting for the photons, three massless neutrino species, electrons and positrons. 

As we previously showed in Eqs. (\ref{eq:bde_rdg_1}) and (\ref{eq:bde_rdg_2}) , the introduction of the DG implies the addition of an extra amount of radiation in the early universe. Using Eq. (\ref{eq:bde_TDGTSM_grl}), the value of $N_{ext}$ in Eq. (\ref{eq:bde_rdg_2}) is:  
\begin{equation}\label{eq:bde_Next}
N_{ext}=\frac{4}{7}g_\textrm{DG}^{gut} \left(\frac{g_\textrm{SM}^{\nu dec}}{g_\textrm{SM}^{gut}}\right)^{4/3}= 0.945, 
\end{equation}
valid for $a_{\nu dec}\leq a < a_c$.  At the phase transition at $\Lambda_c$ (i.e. at $a_c$) the elementary particles of the DG form neutral composite states (i.e. the BDE meson particle is formed) and we no longer have extra relativistic particles, so $N_{ext}=0$ for $a\geq a_c$.
By the time when the condensation occurs at $a_c$, only the photons and neutrinos remain relativistic. The energy density of this standard radiation is given by Eq. (\ref{eq:bde_rhogT_2}), $\rho_r=\tfrac{\pi^2}{30}g_rT_\gamma^4$, where $g_r=2+\tfrac{7}{8}(2)N_\nu \left(T_\nu/T_\gamma\right)^4=3.383$ and $N_\nu=3.046$ accounting for neutrino decoupling effects \cite{Mangano02_Cosmo}. Combining this result with $\rho_\mathrm{DG}=(\pi^2/30)g_\mathrm{DG}T_\mathrm{DG}^4$ and Eq. (\ref{eq:bde_TDGTSM_grl}), we find the ratio of the energy density of the DG to the standard radiation at $a_c$:
\begin{equation}\label{eq:bde_rhoDGrhor_ac}
\frac{\rho_\mathrm{DG}(a_c)}{\rho_{r}(a_c)}=\frac{g_\mathrm{DG}^{gut}}{g_r}\left( \frac{4}{11}\frac{g_\mathrm{SM}^{\nu dec}}{g_\mathrm{SM}^{gut}} \right)^{4/3}=0.1268.
\end{equation}

Since the matter content of the universe is still negligible at that time (see below section \ref{mcmc}), this implies that the DG amounts to the $\Omega_\textrm{DG}(a_c)=\rho_\textrm{DG}/(\rho_r+\rho_\textrm{DG})=0.113$ of the cosmic budget when the condensation happens. For massless neutrinos, the energy density of the standard radiation at $a_c$ is simply related to its present value $\rho_{r0}$ as $\rho_r(a_c)=\rho_{r0}a_c^{-4}$, where $\rho_{r0}$ is proportional to the CMB temperature today $T_\gamma(a_0)=2.7255\mathrm{ K}$ \cite{Fixsen09_CMB}, whereas the energy density of the DG can be expressed as:
\begin{equation}\label{eq:bde_rho_ac}
\rho_\mathrm{DG}(a)=\rho_\mathrm{DG}(a_c) \le(\fr{a_c}{a}\ri)^{4}= 3\Lambda_c^4   \le(\fr{a_c}{a}\ri)^{4},
\end{equation}
where $\rho_\textrm{DG}(a_c) = 3\Lambda_c^4$ (see Eq. (\ref{eq:bde_rhophi_ac})). Therefore, Eq. (\ref{eq:bde_rhoDGrhor_ac}) can be written alternatively as:
\begin{equation}\label{eq:bde_rhoDGrhor_ac_2}
 \frac{\Omega_\mathrm{DG}(a_c)}{\Omega_r(a_c)}= \frac{\rho_\mathrm{DG}(a_c)}{\rho_r(ac)}=\frac{3\Lambda_c^3}{\rho_{r0}a_c^{-4}}=\frac{3(a_c\Lambda_c)^4}{\rho_{r0}},
\end{equation}
and solving for $a_c\Lambda_c$, we arrive at the constraint equation:
\begin{eqnarray}\label{eq:bde_acLc_theory}
\frac{a_c\Lambda_c}{\textrm{eV}} & = & \left(\frac{\rho_{r0}}{3\textrm{eV}^4} \frac{g_\textrm{DG}^{gut}}{g_r} \right)^{1/4} \left(\frac{4}{11} \frac{g_\textrm{SM}^{\nu dec}}{g_\textrm{SM}^{gut}}\right)^{1/3}\nonumber\\
& = &  1.0939 \times 10^{-4},
\end{eqnarray}
which is a meaningful constriction relating the two characteristic quantities of our BDE model, namely, the energy scale $\Lambda_c$ and the scale factor $a_c$ at the condensation epoch. The expansion rate $H \equiv \dot{a}/a$ before the phase transition is given by:
\begin{equation}\label{eq:bde_Friedmann_DG} 
H^2 = \frac{8\pi G}{3}\left[\rho_{m}+\rho_{r}+3(a_c\Lambda_c)^4a^{-4}\right] \;\; \textrm{ for } a<a_c,
\end{equation}
where overdots denote cosmic time derivatives, $\rho_m$ and $\rho_r$ are the energy density of matter and radiation, respectively. 

\subsection{BDE Potential $V(\phi)$}
As the expansion proceeds and the universe cools down, the gauge coupling of the DG interaction becomes strong. The elementary fields of the DG are no longer weakly coupled and they form composite states which interact with the SM sector only through gravity \cite{Macorra03_BDE}. We assume that all the DG particles condense into the lightest state corresponding to a scalar meson $\phi$, which represents the dark energy in our model \cite{Macorra03_BDE}. Therefore, all the energy stored in the DG is completely transferred to our dark energy meson BDE at the moment of the condensation \cite{Macorra03_BDE}: $\rho_\mathrm{DG}(a_c)=\rho_\mathrm{BDE}(a_c)$. 

Strong gauge interactions produce a non-perturbative scalar potential $V$ below $\Lambda_c$. This potential can be computed from the Affleck-Dine-Seiberg superpotential $W=(N_c-N_f)(\Lambda_c^{b_0}/det \langle Q\tilde Q\rangle)^{1/(N_c-N_f)}$ for a non-Abelian $SU(N_c)$ gauge group with $N_f$ massless fields \cite{ADS85_HEP,Burgess97_HEP}. The scalar potential in  global supersymmetry (SUSY) for a canonically normalised meson field $\phi^2\equiv \langle Q\tilde Q\rangle$ is given by $V= |W_\phi|^2$, where $W_\phi=\partial W/\partial \phi$ \cite{Binetruy99_HEP,Masiero99_HEP,*Binetruy00_HEP}, representing  the lightest meson field  corresponding to a pseudo-Goldstone boson. It is worth noticing that the ADS superpotential $W$ is exact (it receives no radiative corrections) and the resulting  scalar potential $V$ is stable against quantum corrections \cite{ADS85_HEP}. The resulting IPL potential at three level is $V(\phi)=\Lambda_c^{4+n}\phi^{-n}$, with \cite{Macorra03_BDE}:  
\begin{equation}\label{eq:bde_V_1}
V(\phi)=\Lambda_c^{4+2/3}\phi^{-2/3}, 
\end{equation}
\begin{equation}\label{eq:bde_V_2}
n=2\left( 1+\frac{2}{N_c-N_f}\right)=\frac{2}{3},
\end{equation}
where the exponent $n$ is a function of $N_c$ and $N_f$, determined only by the number particles of the DG. Global symmetries and SUSY protect the mass of $\phi$, which is given by:
\begin{equation}\label{eq:bde_m_phi}
m_{\phi}^2 \equiv \frac{d^2V}{d\phi^2} = \frac{10}{9} \le( \frac{\Lambda_c}{\phi} \ri)^ {8/3} \Lambda_c^2.
\end{equation}
From dimensional analysis it is natural to expect that all dimensional parameters are to be proportional to the symmetry breaking scale, which in our case is $\Lambda_c$. At the moment of the phase transition denoted by the scale factor $a_c$, we have:
\begin{equation}\label{eq:bde_ic_1}
\phi_c=\Lambda_c,\;\;\; V(\phi_c)=\Lambda_c^4,\;\;\; m_\phi(\phi_c)=\sqrt{\frac{10}{9}} \;\;\Lambda_c
\end{equation}
The dynamical evolution of $\phi$ will be to minimize the potential $V(\phi)$, so we obtain $\phi (a)\geq \Lambda_c$ and $m_\phi(a) \leq  m_\phi(a_c)$. Even thought the potential $V(\phi)$ is exact at $a_c$, a non-vanishing $V$ breaks SUSY and the scalar potential will then receive radiative corrections. These radiative corrections can be determined using the Coleman-Weinberg one-loop effective potential \cite{Coleman73_HEP,Macorra03_BDE,Macorra05_BDE} and are given by:
\begin{eqnarray}\label{eq:bde_V_radiative}
V_1 & = & \frac{1}{32\pi^2}  \int_{0}^{\Lambda_{cut}^2} dp^2 p^2 \ln (p^2+m^2)\nonumber\\
 & = & \frac{\Lambda_{cut}^4}{64\pi^2} \left( x+x^2 \ln \left[\frac{x}{1+x}\right]+\ln[1+x]\right),  
\end{eqnarray}
where $x\equiv m_\phi^2/\Lambda_{cut}^2$. In general the contribution of $V_1$ in Eq. (\ref{eq:bde_V_radiative}) can have strong effects on the evolution of the scalar field, since typically $\Lambda_{cut}$ can be identified with the Planck or the unification scale. However, the cutoff scale in our model is simply $\Lambda_{cut}=\Lambda_c$, because above $\Lambda_c$ the BDE meson particle has not been formed yet. Since the scalar field rolls down the potential from $\phi(a_c)=\Lambda_c$ to $\phi(a_0)\sim m_\textrm{Pl}$ at present, $\Lambda_c \leqslant \phi$ and therefore  $x=(10/9)(\Lambda_c/\phi)^{8/3}\leqslant 1.1$, so the radiative corrections $V =V_o+V_1 \simeq \Lambda_c^{4+2/3} \phi^{-2/3}[1+(\Lambda_c/\phi)^2/(32\pi^2)]$ are negligible at late times and they don't spoil the behaviour of our BDE potential of Eq. (\ref{eq:bde_V_1}). We remark that these results are insensitive to the SUSY breaking scale in the SM. Moreover, all the quantities we have introduced to describe the features of the DG and the scalar potential such as $g_\textrm{DG}^{gut}$, $\Lambda_c$, $a_c$ and $n$ cannot be varied arbitrarily, but they assume a fixed value determined by Eqs. (\ref{eq:bde_gDG}), (\ref{eq:bde_Lc_theory}), (\ref{eq:bde_acLc_theory}) and (\ref{eq:bde_V_2}), respectively. As we stated before, since $N_c$ and $N_f$ are fundamental quantities as in the SM, they are not free cosmological parameters.

\subsection{Cosmological evolution of BDE.}
Once the condensation occurs, the evolution of the light meson particle representing the dark energy is described by a canonical scalar field minimally coupled with the SM sector. The energy density ($\rho_\textrm{BDE}$) and the pressure ($P_\textrm{BDE}$) of the scalar field are given by the generic quintessence expressions \cite{Copeland06_QCDM,*Tsujikawa13_QCDM}:
\begin{equation}\label{eq:bde_rhophi_Pphi}
 \rho_{\textrm{BDE}}= \frac{1}{2}\dot{\phi}^2+V, \;\;\;\;\;\; P_{\textrm{BDE}}=\frac{1}{2}\dot{\phi}^2-V,
\end{equation}
where the self-interaction term $V$ corresponds to the IPL potential of Eq. (\ref{eq:bde_V_1}). Unlike a cosmological constant, where the equation of state of the dark energy (EoS) $w_\Lambda=-1$ is constant in time, here we have a time-varying EoS:
\begin{equation} \label{eq:bde_wphi} 
w_{\textrm{BDE}}\equiv \frac{P_\textrm{BDE}}{\rho_\textrm{BDE}} =\frac{\frac{1}{2}\dot{\phi}^2-V}{\frac{1}{2}\dot{\phi}^2+V},
\end{equation}
whose value at any moment is determined by the competition between the kinetic ($\dot{\phi}^2$) and the potential ($V$) terms. Concordance with the observations imposes the slow-roll condition $\dot{\phi}^2(a_0)\ll V(\phi_0)$ leading to an EoS close to $-1$ at present. The evolution of the scalar field is determined by the Klein-Gordon equation \cite{Copeland06_QCDM,*Tsujikawa13_QCDM}:
\begin{equation} \label{eq:bde_KG} 
\ddot{\phi}+3H\dot{\phi}+\frac{dV}{d\phi} = 0,
\end{equation} 
and now we replace the DG term in Eq. (\ref{eq:bde_Friedmann_DG}) by $\rho_\textrm{BDE}$ to get the Friedmann equation in the presence of the BDE meson:
\begin{equation} \label{eq:bde_Friedmann_phi} 
H^2=\frac{8\pi G}{3}(\rho_{m}+\rho_{r}+\rho_\textrm{BDE}) \hspace*{0.5cm} \textrm{ for } a\geqslant a_c.
\end{equation}
As we stated before in Eq. (\ref{eq:bde_ic_1}), the natural initial condition for the scalar field in our model is $\phi(a_c)=\Lambda_c$, 
since this is just the energy of the symmetry breaking scale in the DG sector corresponding to the formation of bound states, i.e., the BDE meson field \cite{Macorra03_BDE}. The scalar potential at that time is $V(a_c)=\Lambda_c^{4}$ and solving for $\rho_\textrm{BDE}$ and $\dot{\phi}$ in Eqs. (\ref{eq:bde_rhophi_Pphi})  and (\ref{eq:bde_wphi}) we find:
 \begin{equation}\label{eq:bde_phidot_ac}
 \dot{\phi}_c=\sqrt{2\Lambda_c^4\left(\frac{1+w_{\textrm{BDE} c}}{1-w_{\textrm{BDE} c}}\right)}\Bigg|_{w_\textrm{BDEc}=1/3}=2\Lambda_c^2,
 \end{equation}
\begin{equation}\label{eq:bde_rhophi_ac}
\rho_\textrm{BDE}(a_c)=\frac{2\Lambda_c^4}{1-w_\textrm{BDEc}} \bigg|_{w_\textrm{BDEc}=1/3}  =3\Lambda_c^4,
 \end{equation}
where we take $w_\textrm{BDEc}=1/3$ in both expressions, since the DG dilutes as radiation before $a_c$. In any case, we have checked that the evolution of $\rho_\mathrm{BDE}$ and $w_\mathrm{BDE}$ is not sensitive to the chosen value of $w_\mathrm{BDEc}$ (see next section). The crucial point is that in our model the initial conditions of the scalar field can be naturally proposed from physical considerations. Taking the conditions in Eqs. (\ref{eq:bde_ic_1}) and (\ref{eq:bde_phidot_ac}), the amount of dark energy at any time is fully determined by the solution of Eqs. (\ref{eq:bde_KG}) and (\ref{eq:bde_Friedmann_phi}) using Eq. (\ref{eq:bde_Lc_theory}), once  the amount of dark matter $\rho_{m0}$ is given. On the other hand, in the $\Lambda$CDM scenario the amount of dark matter  $\rho_{m0}$ and  dark energy  $\rho_\Lambda=\Lambda/(8\pi G)$---or equivalently, the cosmological constant $\Lambda$---are  free parameters. Consequently, our BDE model has in principle one less free parameter than $\Lambda$CDM. In practice, however,  $\Lambda_c$ is still a not well determined quantity due to the errors in $\Lambda_{gut}$ and $g_{gut}$ of the SM gauge groups. Nevertheless, the theoretical constraint given by Eq. (\ref{eq:bde_Lc_theory}) is definitely more compelling than the discrepancy between the expected and the observed value of $\Lambda$ which spans 30 orders of magnitude.

\begin{figure}\centering \includegraphics[width=1\linewidth, height=0.19\textheight]{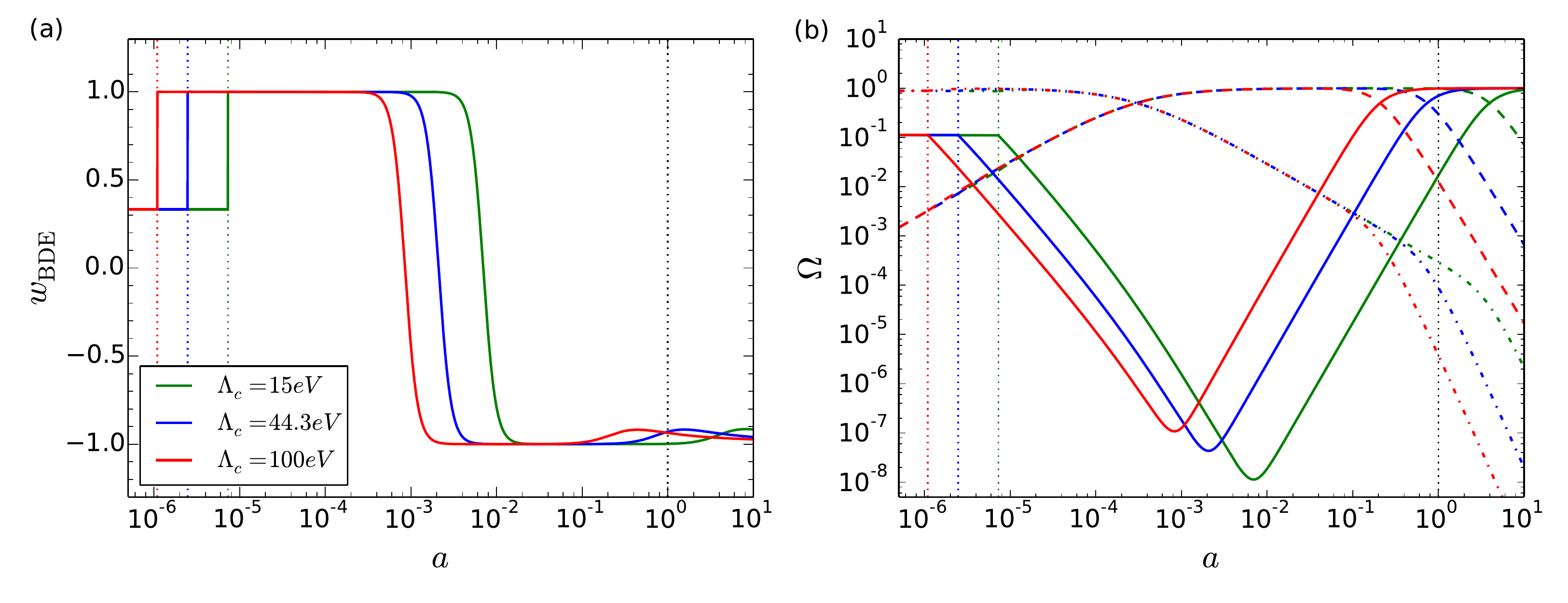}\caption{(a) Evolution of the EoS of the dark energy (firstly as the DG, then as the BDE scalar meson) for three different values of the condensation scale $\Lambda_c$. (b) Evolution of the density parameter of the dark energy (solid), matter (dashed) and radiation (dash-dotted) in each case. The colored vertical lines on the left mark the corresponding epoch of the condensation; the line at $a_0=1$ labels the present time.}
\label{fig:bde_background}
\end{figure}

Figure \ref{fig:bde_background}(a) shows the evolution of the EoS for three different values of the condensation scale. The overall evolution is the same, but the phase transition occurs earlier for larger values of $\Lambda_c$ (c.f. Eq. (\ref{eq:bde_acLc_theory})) and therefore the curves are shifted to the left. Before $a_c$ the DG dilutes as radiation with a constant EoS $w_\textrm{BDE}=1/3$. When the DG particles condense into the BDE meson, the evolution of the EoS is driven by the competition between the kinetic ($\dot{\phi}^2$) and the potential ($V$) terms \cite{Macorra02_BDE}. Initially, the EoS leaps abruptly to $\simeq 1$ and remains at this value for a long period of time. Eventually, the EoS drops to $w_\mathrm{BDE}\simeq -1$ mimicking a cosmological constant. Finally, the EoS departs from $w_\mathrm{BDE} \simeq -1$ forming the small lumps we see at late times and it will be approaching asymptotically to $-1$ in the future. The crucial point is that the dark enery EoS today and its rate of change depend on the initial conditions as shown by the intersection of the curves with the vertical line corresponding to the present epoch.

The evolution of the density parameter $\Omega_i=\rho_i/\rho_{crit}$ with $\rho_{crit}=3H^2/(8\pi G)$ of matter, radiation and BDE is shown in Fig. \ref{fig:bde_background}(b). At high energies $a \ll a_c$, the DG amounts to nearly the 30\% of the energy content of the universe with $\Omega_\textrm{DG}(a_{gut})=0.299$ at the unification scale (see Eq. (\ref{eq:bde_rhoDGrhoSM_gut})). As the relativistic degrees of freedom of the SM decrease, the entropy of the massive particles is transferred to the light ones increasing the temperature of the SM w.r.t. the DG and thus reducing $\Omega_\mathrm{DG}$. For the interval of the scale factor shown in the plot, standard radiation consists only of photons and neutrinos, giving $\Omega_\textrm{DG}(a<a_c) \approx \Omega_\mathrm{DG}(a_c)=0.113$ (c.f. Eq.(\ref{eq:bde_rhoDGrhor_ac_2})). Since the energy of the DG is completely transferred to the BDE meson at the moment of the condensation \cite{Macorra03_BDE}, using Eq. (\ref{eq:bde_rhophi_ac}) we find:
\begin{eqnarray}\label{eq:bde_omegaphi_ac}
\Omega_{\textrm{BDE}}(a_c) &=& \frac{\rho_\mathrm{DG}}{\rho_{r}+\rho_{m}+\rho_\mathrm{DG}}\\
 &=& \frac{3(a_c\Lambda_c)^4}{\rho_{r0}+\rho_{m0}a_c+3(a_c\Lambda_c)^4}.
\end{eqnarray}
This is not a negligible value value of the initial density parameter of the BDE meson. However, since the EoS leaps to $1$ at the condensation epoch, the scalar field dilutes rapidly as $\rho_\textrm{BDE}\propto a^{-6}$ and it becomes subdominant for most of the part of the history of the universe, as expected in any dark energy model. Later on in section \ref{implications} we'll see that the rapid dilution of BDE just after the condensation leaves a distinctive imprint on matter perturbations at small scales. When the EoS drops to $w_\mathrm{BDE} \simeq -1$ and BDE evolves like a cosmological constant, the density parameter of the scalar field grows steadily since matter and radiation keep diluting. Finally, BDE becomes dominant at late times. We see that the amount of dark energy today depends on the condensation scale; low values of $\Lambda_c$ lead to low minimum values of $\Omega_\mathrm{BDE}$, thus delaying the arrive of the dark energy age, while large values of $\Lambda_c$ lead to larger minimum values of $\Omega_\mathrm{BDE}$ leading to a complete dark-energy dominance at present. In this paper we constrain $\Lambda_c$ using cosmological information and see how these bounds are compared with the theoretical range given by Eq. (\ref{eq:bde_Lc_theory}).

The homogeneous picture of the universe valid at large scales must be refined to account for all the structure we observe today. Unlike the $\Lambda$CDM scenario where the dark energy is completely homogeneous, consistency with the equivalence principle \cite{Caldwell98_QCDM} means that both the DG and the scalar field must fluctuate in response to matter and radiation inhomogeneities. In this paper we study the linear dynamics, where the size of the fluctuations is small enough to be described accurately by linear perturbation theory \cite{MaBertschinger95_Cosmo}. We assume that after neutrino decoupling, when we start solving the equations, the DG perturbations behave as the neutrino perturbations, since both components interact with the other particles only through gravity. After the condensation, we decompose the scalar field into a sum of a homogeneous term $\bar{\phi}(\eta)$ and a position-dependent perturbation $\delta\phi(\eta,\textbf{x})$, where $d\eta\equiv dt/a$ is the conformal time and $\textbf{x}$ is the position in comoving coordinates. Working in the CDM synchronous gauge \cite{MaBertschinger95_Cosmo} defined by the line element $ds^2=a^2(\eta)[-d\eta^2+(\delta_{ij}+h_{ij})dx^idx^j]$, the perturbations of the energy density $\delta \rho\equiv \rho(\eta,\textbf{x})-\bar{\rho}(\eta)$ and the pressure $\delta P\equiv P(\eta,\textbf{x})-\bar{P}(\eta)$ of the scalar field are given by \cite{Caldwell98_QCDM}:
\begin{equation} \label{eq:bde_deltarho_deltaP_1} 
\delta\rho_\mathrm{BDE} = \frac{\bar{\phi}'\delta \phi'}{a^2}+V_\phi\delta \phi, 
\end{equation}
\begin{equation}\label{eq:bde_deltarho_deltaP_2} 
\delta P_\mathrm{BDE} = \frac{\bar{\phi}'\delta \phi'}{a^2}-V_\phi \delta \phi,
\end{equation}  
where the primes stand for conformal time derivatives and $V_\phi \equiv dV/d\phi$. The evolution of $\delta \phi$ in Fourier space is determined by \cite{Caldwell98_QCDM}:
\begin{equation}\label{eq:bde_KG_perturbed}
 \delta\phi''+2\mathcal{H}\delta\phi'+(k^2+a^2V_{\phi\phi})\delta\phi=-\frac{1}{2}\bar{\phi}'h',
\end{equation}
where $k$ is the Fourier mode, $\mathcal{H}\equiv a'/a$ is the conformal expansion rate and $h=\textrm{Tr}(h_{ij})$ ---not to be confused with the adimensional Hubble constant $h\equiv H_0/100$. The perturbations of the scalar field couple with the other fluids through $h$, which is directly proportional to the CDM overdensities $\delta_c \equiv \delta \rho_c/\bar{\rho}_c$ in this gauge \cite{MaBertschinger95_Cosmo}. The initial conditions for $\delta \phi$ and $\delta \phi'$ are provided by matching the DG and the scalar field overdensities $\delta_\textrm{BDE}=\delta \rho_\textrm{BDE}/\bar{\rho}_\textrm{BDE}$ at $a_c$. However, it has been shown \cite{Brax00_QCDM} that for IPL potentials such as ours, the solution of Eq. (\ref{eq:bde_KG_perturbed}) is insensitive to them. Based on this result we take $\delta \phi'(a_c)=0$, and using Eqs. (\ref{eq:bde_ic_1}) and (\ref{eq:bde_deltarho_deltaP_1}) we get $\delta \phi (a_c)=-\tfrac{9}{2}\Lambda_c\delta_\textrm{DG}(a_c)$. BDE perturbations don't grow with time, but they fluctuate around the homogeneous background with a non-constant but a very small amplitude. Nevertheless, although the dark energy is very smooth also in our model, the small inhomogeneities do have an effect on the CMB and the evolution of matter perturbations as we'll discuss in section \ref{implications}.

\section{Cosmological constraints.}\label{mcmc}
\begin{table*}\caption{\small{Best fit, mean and 68\% confidence limits for the BDE and $\Lambda$CDM models from the joint analysis of the CMB temperature anisotropy spectrum data \cite{PlanckCP15_CMB}, SNeIa distance measurements \cite{Betoule14_SNeIa} and BAO information \cite{GilMarin16_BAO,Ross15_BAO,Beutler11_BAO}. The last two rows show the goodness of the fit ($\chi^2$) decomposed into the contribution from each probe and the prior for the best fit models.}\normalsize{}}
 \centering
{\setlength{\extrarowheight}{2.5pt}
\begin{ruledtabular}
\begin{tabular}{l c c c c}
\multirow{3}{*}{Parameter} & \multicolumn{2}{c}{BDE} & \multicolumn{2}{c}{$\Lambda$CDM}\\\cline{2-5}
& Best fit & 68\% limits & Best fit & 68\% limits \\\hline
\footnotesize{$\Lambda_c$ [eV]} & \footnotesize{44.02} & \footnotesize{44.09 $\pm$ 0.28} & \footnotesize{...} & \footnotesize{...}\\
\footnotesize{$10^{6}a_c$} & \footnotesize{2.48} & \footnotesize{2.48 $\pm$ 0.02} & \footnotesize{...} & \footnotesize{...} \\
\footnotesize{$\Omega_\textrm{BDE}(a_c)$} & \footnotesize{0.1117} & \footnotesize{0.11174 $\pm$ 0.00001} & \footnotesize{...} & \footnotesize{...}\\
\hline
\footnotesize{$\Omega_b h^2$} & \footnotesize{0.02252} & \footnotesize{0.02257 $\pm$ 0.00021} & \footnotesize{0.02243} & \footnotesize{0.02238 $
\pm$ 0.00021}\\
\footnotesize{$\Omega_c h^2$} & \footnotesize{0.1173} & \footnotesize{0.1171 $\pm$ 0.0013} & \footnotesize{0.1181} & \footnotesize{0.1182 $\pm$ 0.0012}\\
\footnotesize{$100\theta_{\textrm{MC}}$} & \footnotesize{1.04106} & \footnotesize{1.04112$\pm$ 0.00043} & \footnotesize{1.04113} & \footnotesize{1.04112 $\pm$ 0.00042}\\
\footnotesize{$\ln (10^{10}A_s)$} & \footnotesize{3.164} & \footnotesize{3.179$^{+0.055}_{-0.048}$} & \footnotesize{3.165} & \footnotesize{3.150 $\pm$ 0.052}\\
\footnotesize{$n_s$} & \footnotesize{0.9774} & \footnotesize{0.9780 $\pm$ 0.0050} & \footnotesize{0.9710} & \footnotesize{0.9701 $\pm$ 0.0049}\\
\footnotesize{$\tau$} & \footnotesize{0.117} & \footnotesize{0.124$^{+0.028}_{-0.025}$} & \footnotesize{0.118} & \footnotesize{0.110 $\pm$ 0.027}\\\hline
\footnotesize{$H_0$ [km s$^{-1}$Mpc$^{-1}$]} & \footnotesize{67.68} & \footnotesize{67.82 $\pm$ 0.55} & \footnotesize{68.64} & \footnotesize{68.57 $\pm$ 0.58}\\
\footnotesize{$\Omega_\textrm{DE0}$} & \footnotesize{0.695} & \footnotesize{0.696 $\pm$ 0.007} & \footnotesize{0.702} & \footnotesize{0.701 $\pm$ 0.007}\\
\footnotesize{$\Omega_{m}$} & \footnotesize{0.305} & \footnotesize{0.304 $\pm$ 0.007} & \footnotesize{0.298} & \footnotesize{0.299 $\pm$ 0.007}\\
\footnotesize{$w_\textrm{DE0}$} & \footnotesize{-0.9296} & \footnotesize{-0.9294 $\pm$ 0.0007} & \footnotesize{$-1$} & \footnotesize{$-1$}\\
\footnotesize{$z_{*}$} & \footnotesize{1089.98} & \footnotesize{1089.90 $\pm$ 0.32} & \footnotesize{1089.68} & \footnotesize{1089.75 $\pm$ 0.32}\\
\footnotesize{$r_*$ [Mpc]} & \footnotesize{144.89} & \footnotesize{144.91 $\pm$ 0.32} & \footnotesize{144.88} & \footnotesize{144.89 $\pm$ 0.31}\\
\footnotesize{$D_{\rm{A}}(z_{*})$ [Gpc]} & \footnotesize{13.92} & \footnotesize{13.92 $\pm$ 0.03} & \footnotesize{13.91} & \footnotesize{13.92 $\pm$ 0.03}\\
\footnotesize{$r_{\rm{drag}}$ [Mpc]} & \footnotesize{147.52} & \footnotesize{147.54 $\pm$ 0.34} & \footnotesize{147.53} & \footnotesize{147.56 $\pm$ 0.34}\\
\footnotesize{$z_{eq}$} & \footnotesize{3342} & \footnotesize{3338 $\pm$ 29} & \footnotesize{3359} & \footnotesize{3360 $\pm$ 29}\\
\footnotesize{$k_{D}$ [Mpc$^{-1}$]} & \footnotesize{0.1400} & \footnotesize{0.13998 $\pm$ 0.00045} & \footnotesize{0.1404} & \footnotesize{0.14037 $\pm$ 0.00045}\\
\footnotesize{$Y_P^{\rm{BBN}}$} & \footnotesize{0.2588} & \footnotesize{0.2588 $\pm$ 0.0001} & \footnotesize{0.2467} & \footnotesize{0.2467 $\pm$ 0.0001}\\
\footnotesize{$10^5D/H$} & \footnotesize{2.89} & \footnotesize{2.88 $\pm$ 0.04} & \footnotesize{2.58} & \footnotesize{2.59 $\pm$ 0.04}\\\hline
\multicolumn{5}{l}{\footnotesize{$\chi^2(\rm BDE)$ = 5.609\textrm{ (BAO)} + 776.510\textrm{ (CMB)} + 695.668\textrm{ (SNeIa)} + 1.833\textrm{ (prior)} = 1479.621}} \\
\multicolumn{5}{l}{\footnotesize{$\chi^2(\Lambda \rm CDM)$ = 7.115\textrm{ (BAO)} + 776.884\textrm{ (CMB)} + 695.075\textrm{ (SNeIa)} + 1.681\textrm{ (prior)} = 1480.754}}
\label{tab:mcmc_table} \end{tabular} \end{ruledtabular}} 
\end{table*}

The distinctive dynamics of the dark energy in our BDE model leaves imprints on cosmological quantities that can be probed by current observational data. In order to find the constraints on the parameters consistent with the observations, we perform a Markov-Chain Monte Carlo (MCMC) analysis using the \texttt{CAMB} \cite{Lewis00_CAMB} and \texttt{CosmoMC} \cite{Lewis02_CosmoMC} codes properly adapted to the BDE scenario described above. Our analysis combines the Planck 2015 measurements \cite{PlanckCP15_CMB} of the CMB temperature anisotropy spectrum in the multipole range $l=2-2508$, the JLA compilation \cite{Betoule14_SNeIa} of the luminosity distance of 740 identified type Ia supernovae ranging between $z_{min}=0.01$ and $z_{max}=1.3$, and the BAO signal from three galaxy surveys: the Main Galaxy Sample (MGS) at $z_{eff}=0.15$ \cite{Ross15_BAO}, the 6dF Galaxy Survey (6dF) at $z_{eff}=0.106$ \cite{Beutler11_BAO} and the LOWZ and CMASS samples of the BOSS-DR12 survey \cite{GilMarin16_BAO} at $z_{eff}=0.32$ and $z_{eff}=0.57$, respectively.

\begin{figure}[b]\centering \includegraphics[width=1.\linewidth, height=0.16\textheight]{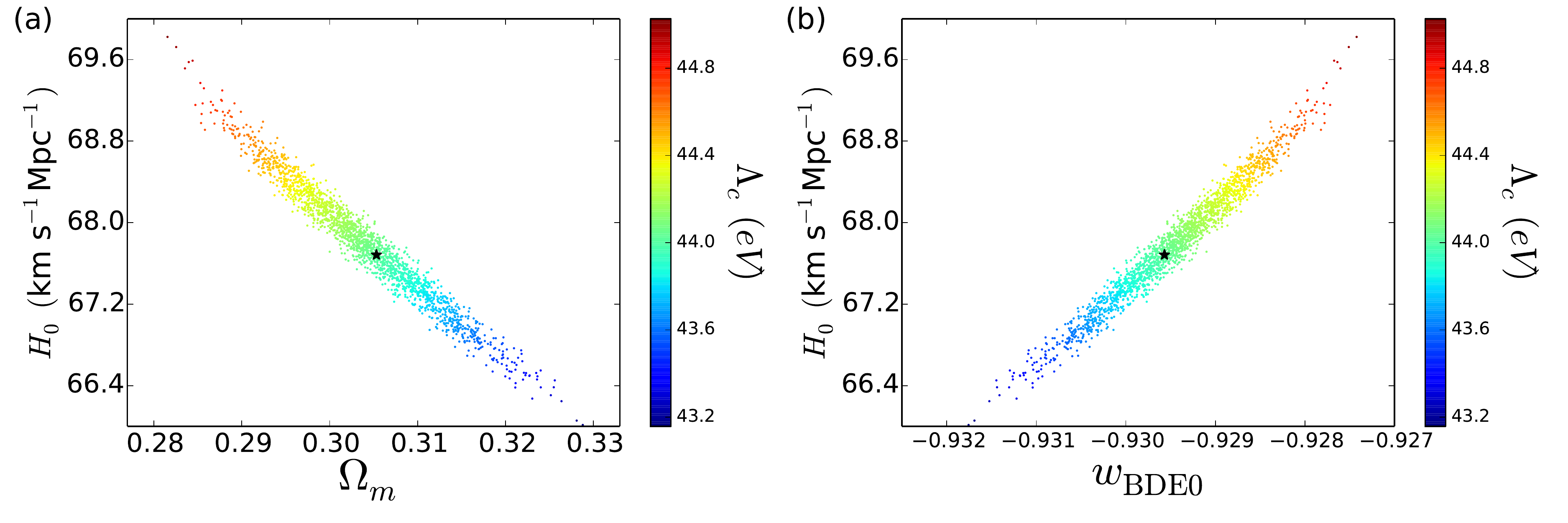}\caption{(a) Samples in the $\Omega_m-H_0$ plane colored by the condensation scale $\Lambda_c$. (b) Samples in the $w_\textrm{BDE0}-H_0$ plane. The stars mark the location of the best fit point.}\label{fig:mcmc_bde_lc_bar}
\end{figure}

\begin{figure*}\centering                  
\includegraphics[width=1.\linewidth, height=0.21\textheight]{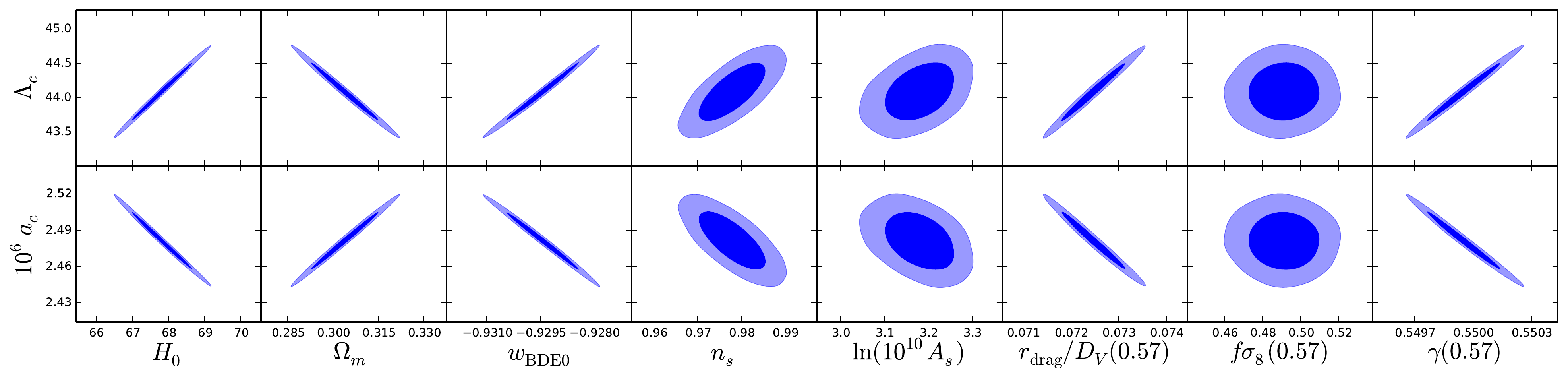}\caption{Joint constraints at 68\% and 95\% confidence limits of the condensation scale $\Lambda_c$ (in eV) and the condensation epoch $a_c$ of the BDE meson with some parameters.}\label{fig:mcmc_bde_rectangle}
\end{figure*}

The set of basic parameters in our dark energy model does not include $H_0$, since this quantity is determined by the solution of the dynamical system of equations for the background, which in turn depends on $\Lambda_c$ through the initial conditions, as we saw before. However, we replace $H_0$ by $\Lambda_c$ as a basic parameter in our MCMC runs, for although $\Lambda_c$ is not a free parameter, we recall that it is still poorly measured because of the uncertainties of the high-energy physics quantities determining its value \cite{Bourilkov15_HEP}. Therefore, the set of basic parameters in our MCMC runs is the same size both in BDE and $\Lambda$CDM. We explore the parameter space with 8 Markov Chains long enough to probe the tails of the posteriors ensuring the convergence of the chains after a burn-in period of 30\% \cite{Trotta17_Statistics}. We also find the best fit point by running several times the corresponding minimisation routines from different initial positions in the parameter space \cite{Trotta17_Statistics}. For BDE, we allow $\Lambda_c$ to vary freely assuming a flat prior between $10^{-3}$ eV and $10^{3}$ eV and we get $a_c$ from Eq. (\ref{eq:bde_acLc_theory}).\\\indent
Table \ref{tab:mcmc_table} lists the mean with the 68\% CL of the marginalised distributions as well as the best fit values of some selected parameters for both models. The bounds on $\Lambda_c$ are consistent with the theoretical estimation of Eq. (\ref{eq:bde_Lc_theory}). This is a strinking difference w.r.t. $\Lambda$CDM, where the observational constraints on the cosmological constant $\Lambda$ differ from the theoretical estimations by 30 orders of magnitude (assuming that the dark energy corresponds to the vacuum state using the Planck mass as a cutoff scale) \cite{Weinberg89_CC,Martin12_CC}. In our BDE model, on the other hand, there is a remarkable agreement between theory and observations on the value of $\Lambda_c$. Moreover, we note that cosmological data impose tighter constraints as seen in the narrower marginal limits in Table \ref{tab:mcmc_table}. Figure \ref{fig:mcmc_bde_lc_bar} shows the relation between $\Lambda_c$ and the joint constraints of expansion rate ($H_0$) with the matter density parameter ($\Omega_m$) and the EoS ($w_\textrm{BDE0}$) at present time. Larger values of $\Lambda_c$ lead to a larger expansion rate and a larger dark energy EoS, while the density of matter of the universe decreases. This is also seen in Fig. \ref{fig:mcmc_bde_rectangle}, where we plot the 68\% and 95\% confidence contours of $\Lambda_c$ and $a_c$ with these three parameters as well as the spectral scalar index ($n_s$), the amplitude ($A_s$) of the primordial spectrum and other quantities describing structure formation (see next section). In some cases we find marked degeneracies as shown by the thinnest contours. The opposite orientation between the top and the bottom row contours simply arises because of the theoretical constriction Eq.  (\ref{eq:bde_acLc_theory}) relating $\Lambda_c$ and $a_c$.

\begin{figure}[b]\centering \includegraphics[width=1.\linewidth]{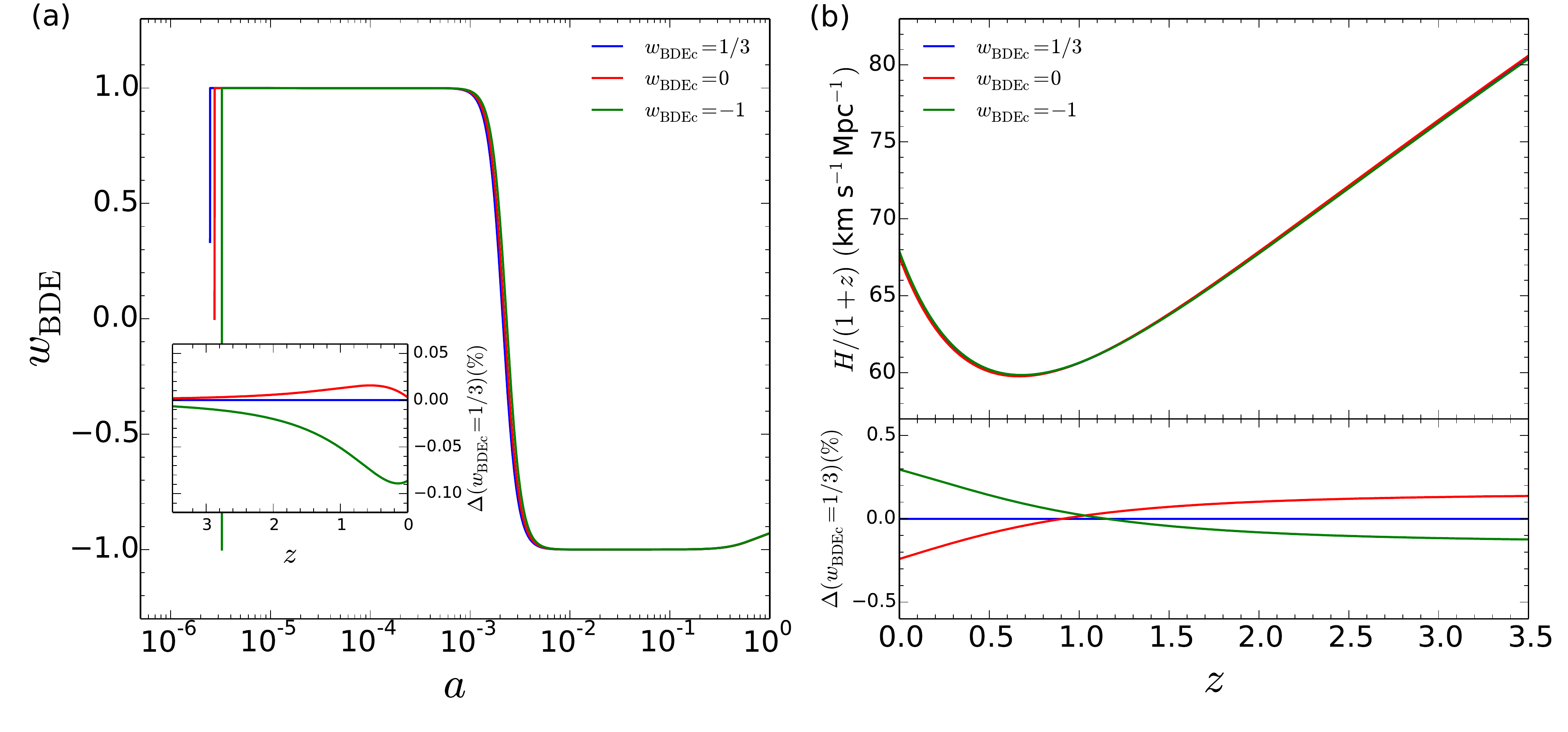}\caption{(a) Evolution of the EoS in BDE for the best fit for three models with different initial EoS: $w_\mathrm{BDEc}=\tfrac{1}{3},0,-1$. The inner subplot shows the relative difference (in \%) w.r.t. the $w_\mathrm{BDEc}=\tfrac{1}{3}$ curve at late times. (b) Conformal expansion rate for each model at late times. The bottom panel shows the relative difference (in \%) w.r.t. the $w_\mathrm{BDEc}=\tfrac{1}{3}$ curve.}\label{fig:implications_wbdec}
\end{figure}

We test the robustness of our results by considering other options such as $w_\mathrm{BDEc}=0$ and $w_\mathrm{BDEc}=-1$ as the initial value of the EoS of the BDE meson (see Eq. (\ref{eq:bde_rhophi_ac})). In each case, we find the best fit point and compute the difference w.r.t. our results with $w_\mathrm{BDEc}=1/3$. Figure \ref{fig:implications_wbdec} shows the evolution of the EoS and the expansion rate. We see that our results are quite insensitive to the chosen value of $w_\textrm{BDEc}$. The differences at late times w.r.t. $w_\mathrm{BDEc}=1/3$ are very small, lying below $0.09\%$ for the EoS and $0.3\%$ for $H$, respectively. Therefore, the late-time dynamics of the dark energy and its imprints on the cosmological distances remain unaltered (see section \ref{implications}). Additionaly, we test the theoretical constriction Eq. (\ref{eq:bde_acLc_theory}) by letting $\Lambda_c$ and $a_c$ vary freely and independently. In this case, we find the best fit values $\Lambda_c=44.06$ eV and $a_c=2.48\times 10^{-6}$ leading to $a_c\Lambda_c/\textrm{eV}=1.0916\times 10^{-4}$, which remarkably deviates from Eq. (\ref{eq:bde_acLc_theory}) only by 0.2\%.

\begin{figure}\centering \includegraphics[width=1.\linewidth]{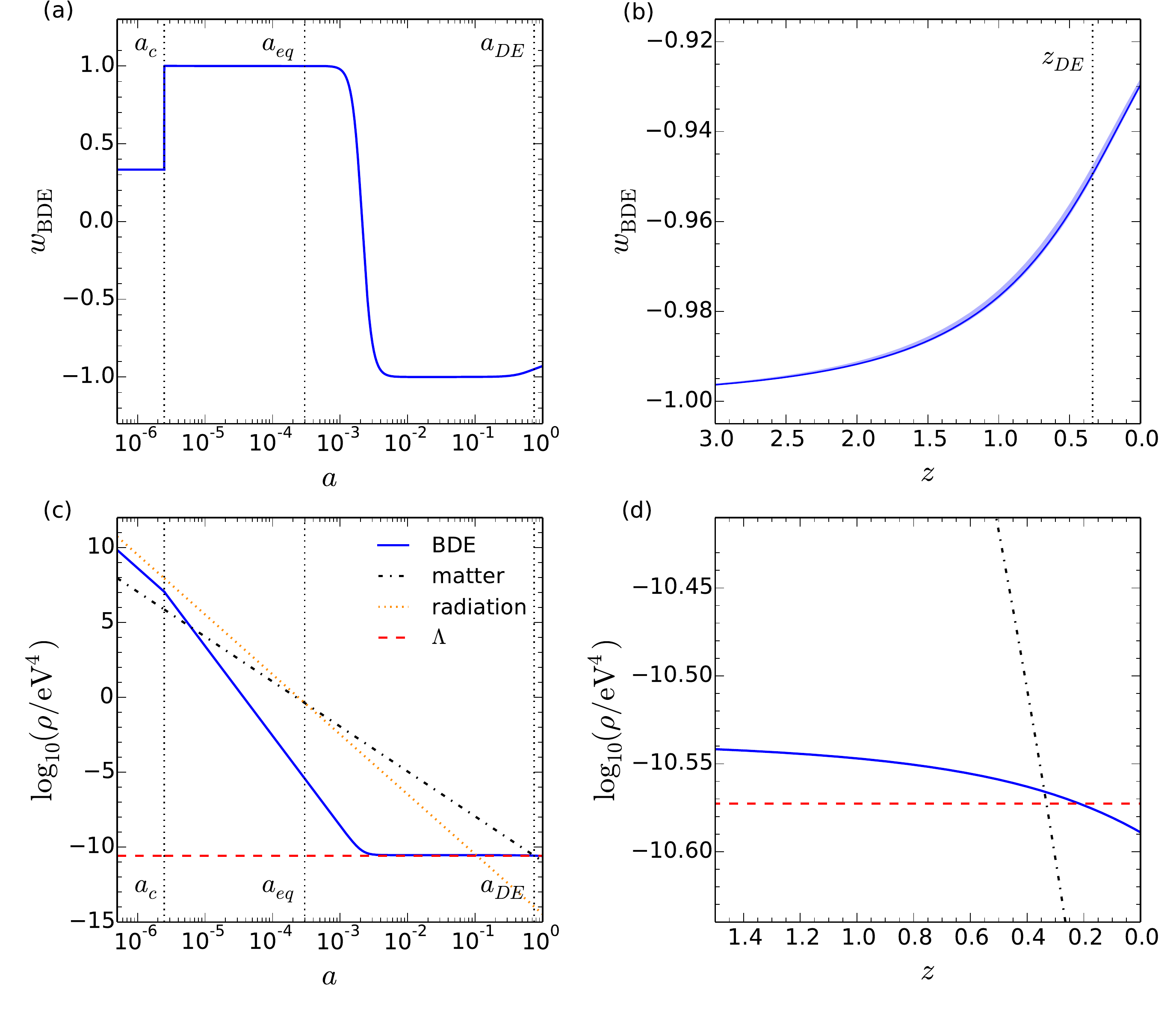}\caption{(a) Evolution of the EoS in BDE for the best fit. The vertical lines mark the condensation epoch ($a_c$), the matter-radiation ($a_{eq}$) and the matter-dark energy equality eras ($a_{DE}$). (b) EoS in BDE at late times. The blue band shows the 95\% confidence ranges obtained from our MCMC samples; $1+z_{DE}=a_{DE}^{-1}$. (c) Dilution of the energy density of matter, radiation, dark energy  in BDE (firstly as the DG, then as the BDE scalar meson) and dark energy as a cosmological constant $\Lambda$. (d) Dilution of the energy density of matter and dark energy at late times.}\label{fig:mcmc_bde_background}
\end{figure}

The overall evolution of the EoS of BDE for the best fit model is plotted in Fig. \ref{fig:mcmc_bde_background}(a). We observe the general features we described previously in Fig. \ref{fig:bde_background}(a), but in this case $w_\textrm{BDE0}$ is located just before reaching the top of the lump. Figure \ref{fig:mcmc_bde_background}(b) zooms in on late times, where the EoS grows monotonically from $-1$ towards its present value. From the thickness of the blue band we see that the EoS is very tightly constrained, allowing us to get accurate bounds from the best fit curve alone. The EoS lies in the range $-1< w_\textrm{BDE}\leqslant -0.999,-0.950$ for $132>z \geqslant 1.8,0.35$, respectively, reaching the mid point between $-1$ and $w_\textrm{BDE0}$ at $z_{mid}=0.65$ with $dw/dz=-1.26$. We provide a useful fitting formula for the EoS in the appendix. Fig. \ref{fig:mcmc_bde_background}(c) shows the evolution of the energy density of matter, standard radiation (i.e., photons and neutrinos) and BDE for the best fit model. We also plot $\rho_\Lambda$, which clearly illustrates the naturalness problems of the cosmological constant. As we have seen, the DG amounts to a non-negligible fraction of the energy content of the early universe, ranging from 30\% at the unification scale to 11\% at $a_c$ (c.f. Eqs. (\ref{eq:bde_rhoDGrhoSM_gut}) and (\ref{eq:bde_rhoDGrhor_ac})). The narrow CL of $\Omega_\textrm{BDE}(a_c)$ in Table \ref{tab:mcmc_table} is the effect of the tiny contribution of $\rho_m$ in Eq. (\ref{eq:bde_omegaphi_ac}). The DG particles condense into BDE almost 5 $e-$folds before the matter-radiation equality epoch when the photon temperature is $T_\gamma(a_c)=2.7255K/a_c=94.7\mathrm{ eV}$ and the density of radiation and matter is $\Omega_{r}(a_c)=0.881$ and $\Omega_{m}(a_c)=0.007$, respectively. However, the stiff behaviour $w_\textrm{BDE}=1$ of the EoS just after $a_c$ leads to the rapid dilution of the scalar field, which is completely halted when the EoS drops to $\approx -1$ and $\rho_\textrm{BDE}$ is frozen evolving nearly as a cosmological constant. The density parameter of BDE at its minimum value is $\Omega_\mathrm{BDE}(z=477)=4.3\times 10^{-8}$. From this time onwards $\Omega_\mathrm{BDE}$ grows, but it is still subdominant until recent times when matter has finally diluted enough. For the best fit we obtain $\Omega_\textrm{BDE}\leqslant 1\%,0.1\%$ for $z \geqslant 5.3,12.7$, respectively and the matter-dark energy equality epoch is reached at $z_{DE}=0.34$. Finally, BDE dilutes further at late times and this extra dilution is what makes $\rho_\textrm{BDE} (a_0)<\rho_\Lambda(a_0)$ by 3.7\% according to the data of Table \ref{tab:mcmc_table} and Fig. \ref{fig:mcmc_bde_background}(d). On the other hand, the difference on the total amount of matter today $\rho_{m0}\propto \Omega_mh^2$ is very small with a tiny excess of 0.5\% in $\Lambda$CDM, leading to an earlier matter-radiation equality epoch $z_{eq}(\textrm{BDE})<z_{eq}(\Lambda\textrm{CDM})$. Consequently, the expansion rate at present time $H_0$ is also larger in $\Lambda$CDM.

\begin{figure*}\centering \includegraphics[width=1.\linewidth]{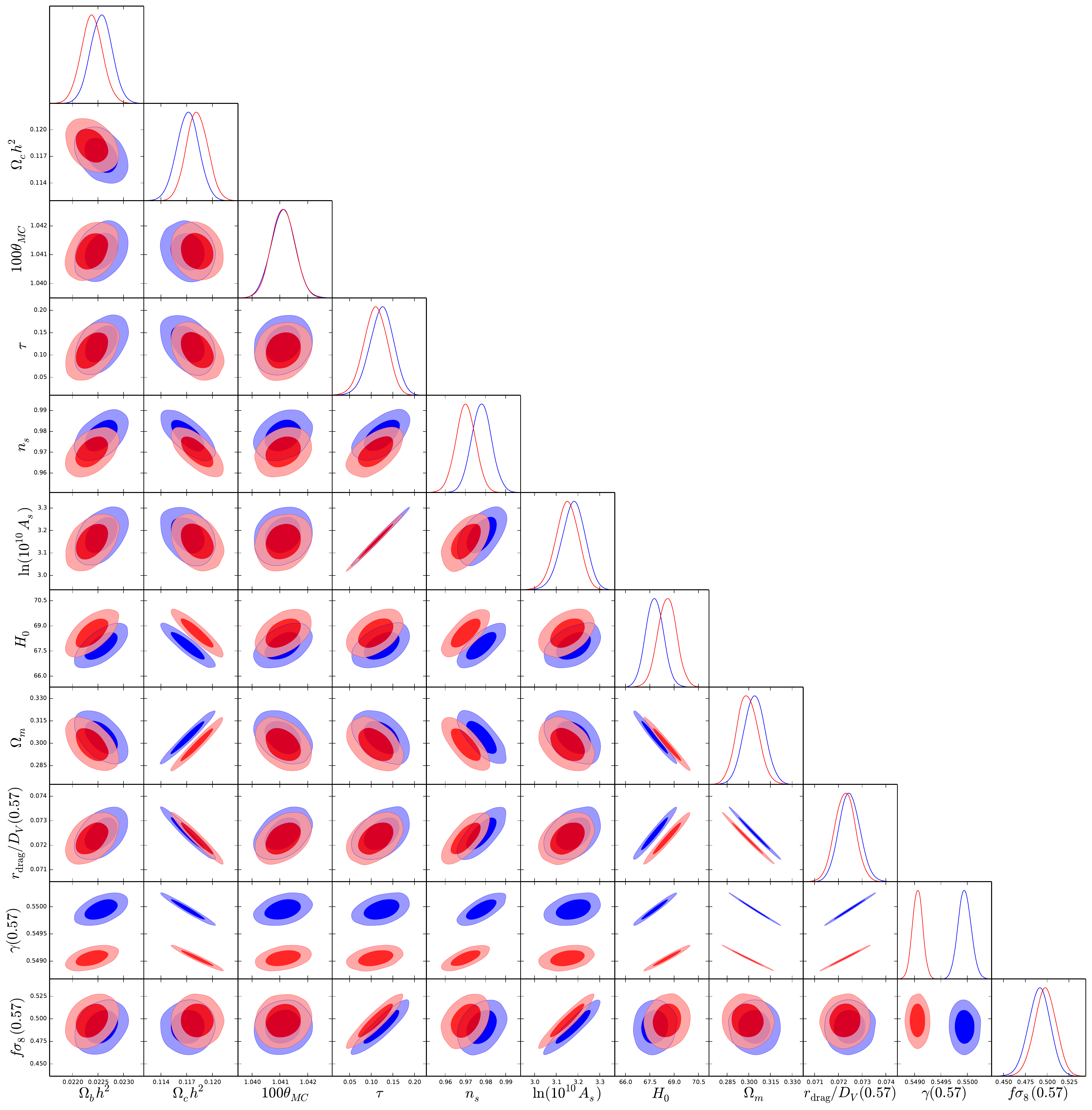}\caption{Marginalised distributions and 68\% and 95\% confidence contours of some cosmological parameters for the BDE (blue) and $\Lambda$CDM (red) models. Our analysis combines CMB temperature anisotropy spectrum data \cite{PlanckCP15_CMB}, SNeIa distance measurements \cite{Betoule14_SNeIa} and BAO information \cite{GilMarin16_BAO,Ross15_BAO,Beutler11_BAO}.}\label{fig:mcmc_triangle}
\end{figure*}

The constraints on the base $\Lambda$CDM parameters \cite{PlanckCP13_CMB,PlanckCP15_CMB} ($\Omega_bh^2$, $\Omega_ch^2$, $100\theta_\mathrm{MC}$, $\tau$, $\ln [10^{10}A_s]$, and $n_s$) agree between the two models within the $1\sigma$ level, as shown in the data of Table \ref{tab:mcmc_table} and Fig. \ref{fig:mcmc_triangle}. However, since the set of basic parameters in both models is the same size and $\chi^2(\rm BDE)<\chi^2(\Lambda \rm CDM)$, BDE fits better the data \cite{Liddle04_Statistics}. Looking at the decomposition of $\chi^2$ in Table \ref{tab:mcmc_table} we see that the decisive contribution comes from the BAO datasets, where BDE significantly improves the likelihood by a ratio of 2.1 w.r.t. $\Lambda$CDM corresponding to a $\chi^2_\textrm{BAO}$ reduction of 21\%. 

\begin{figure}[t]\centering                  
\includegraphics[width=1.\linewidth]{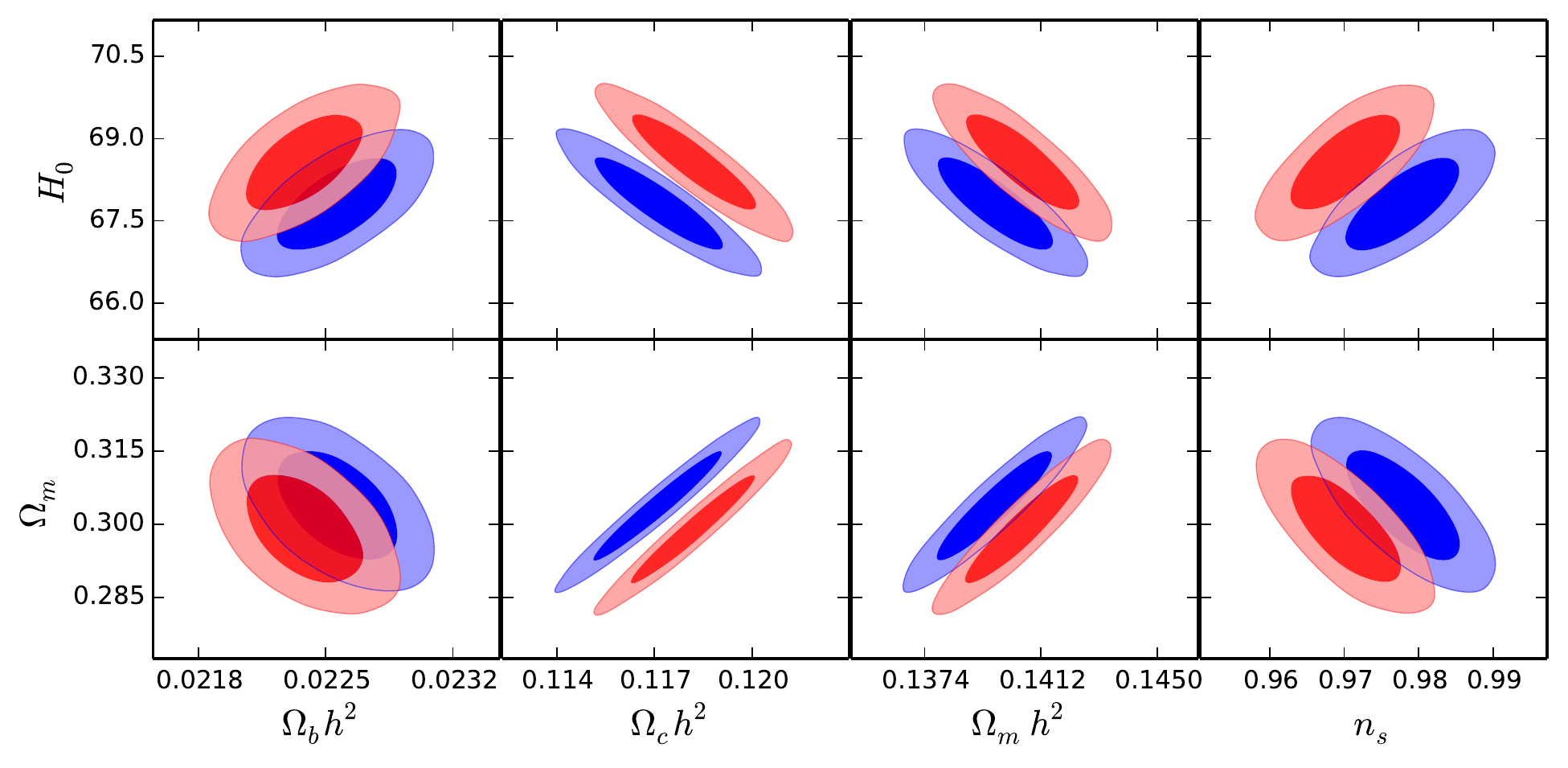}\caption{Joint constraints on $H_0$ and $\Omega_m$ with the present amount of baryons ($\Omega_bh^2$), CDM ($\Omega_ch^2$), baryons + CDM ($\Omega_mh^2$) and the spectral scalar index ($n_s$) for BDE (blue) and $\Lambda$CDM (red). The contours cover the 68\% and 95\% confidence regions.}\label{fig:mcmc_contours}
\end{figure}

However, we find interesting tensions between BDE and $\Lambda$CDM once the amount of dark energy becomes relevant. These tensions can be seen in distance measurements such as the BAO ratio $r_\mathrm{BAO}(z)=r_\mathrm{drag}/D_V(z)$ at different redshifts and structure formation parametrized at late times by the growth index $\gamma$. The discrepancies are clearly seen in the contours of Fig. (\ref{fig:mcmc_triangle}) and we discuss them in detail in the next section. In view of the importance of BAO measurements and LSS data within the forthcoming years \cite{DESI,LSST,Euclid}, these results may provide key evidence for elucidating the nature of the dark energy. Additional interesting tensions  w.r.t $\Lambda$CDM arise when we consider the difference in the content of matter and dark energy at present time. Figure (\ref{fig:mcmc_contours}) shows the joint constraints on $H_0$ and $\Omega_m$ with other patameters. Although there are more baryons ($\Omega_bh^2$) in BDE, the tiny excess of matter in $\Lambda$CDM is due to its larger amount of cold dark matter ($\Omega_ch^2$). The contours overlap in the $\Omega_mh^2-H_0$ and $\Omega_mh^2-\Omega_m$ planes. However, although the contribution from the baryons to $\Omega_mh^2$ is relatively small, if we subtract the larger amount of baryons in BDE the contours split apart as in $\Omega_ch^2-H_0$ and $\Omega_ch^2-\Omega_m$, leading to marked tensions between the models. In the top row the $\Lambda$CDM contours lie above the BDE ones simply because $H_0(\Lambda\textrm{CDM})>H_0(\textrm{BDE})$. If we now consider the matter density parameter $\Omega_m \propto \rho_{m0}/H_0^2$, this result implies that $\Omega_m(\Lambda\textrm{CDM})<\Omega_m(\textrm{BDE})$ and now the BDE contours lie above the $\Lambda$CDM ones as the bottom row. Finally, although they are still in agreement within the $2\sigma$ level, we find modest tensions in contours involving the scalar spectral index. 

\section{Cosmological implications of the model.}\label{implications}
\subsection{The expansion rate and cosmological distances.}
\begin{figure}[b]\centering \includegraphics[width=1.\linewidth]{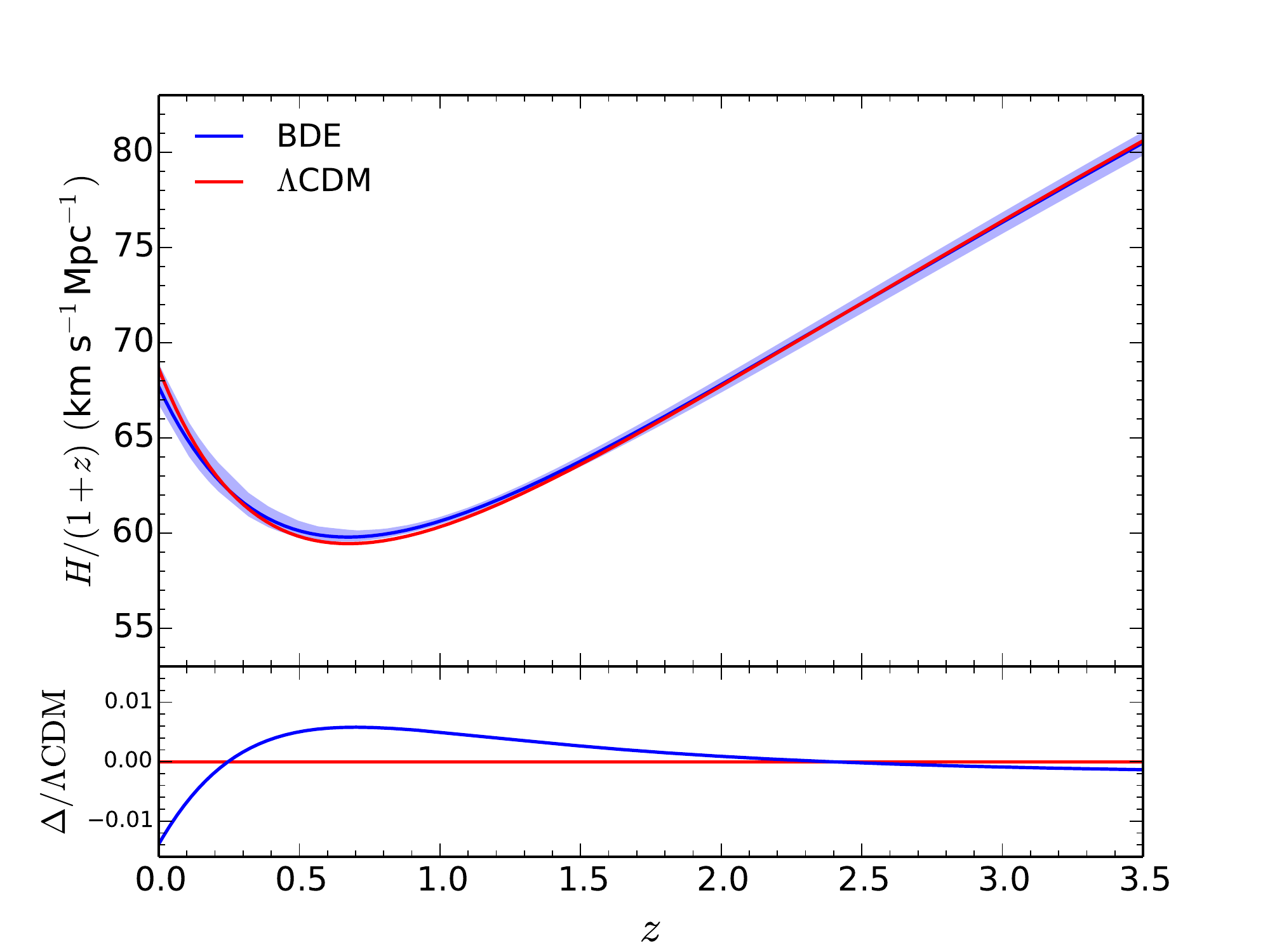}\caption{Conformal expansion rate in BDE (blue) and $\Lambda$CDM (red) at late times. The solid lines correspond to the best fit models; the blue band shows the 95\% confidence regions in BDE obtained from our MCMC samples. The bottom panel displays the fractional difference $\Delta H/H$ of the best fit w.r.t. $\Lambda$CDM.}\label{fig:implications_h}
\end{figure}
The immediate consequence of considering a time-varying equation of state of the dark energy is the modification of the expansion rate of the universe. Figure \ref{fig:implications_h} shows the evolution of the conformal expansion rate $H(z)/(1+z)$ at late times ---when the radiation content of the universe can be neglected--- for the best fit BDE and $\Lambda$CDM models. As we have seen, the density of matter today ($\rho_{m0}$) is roughly the same, while the difference in dark energy is about 3.7\%. This difference is what makes the expansion rate in BDE slower for $z\leqslant 0.24$. However, since $w_\textrm{BDE}$ deviates from $-1$ at late times, $\rho_\textrm{BDE}$ grows faster than $\rho_\Lambda$ as we move to higher redshifts and therefore the expansion rate is larger in BDE  between $0.24 \leqslant z \leqslant 2.4$, leading to a maximum deviation of 0.6\%  w.r.t. $\Lambda$CDM at $z\approx 0.7$. For higher redshifts in the matter domination era the difference in the dark energy content is irrelevant and it is only the tiny difference of 0.5\% in $\rho_{m0}$ what makes the expansion rate slower in BDE once again. 
\begin{figure}[t]\centering \includegraphics[width=1\linewidth]{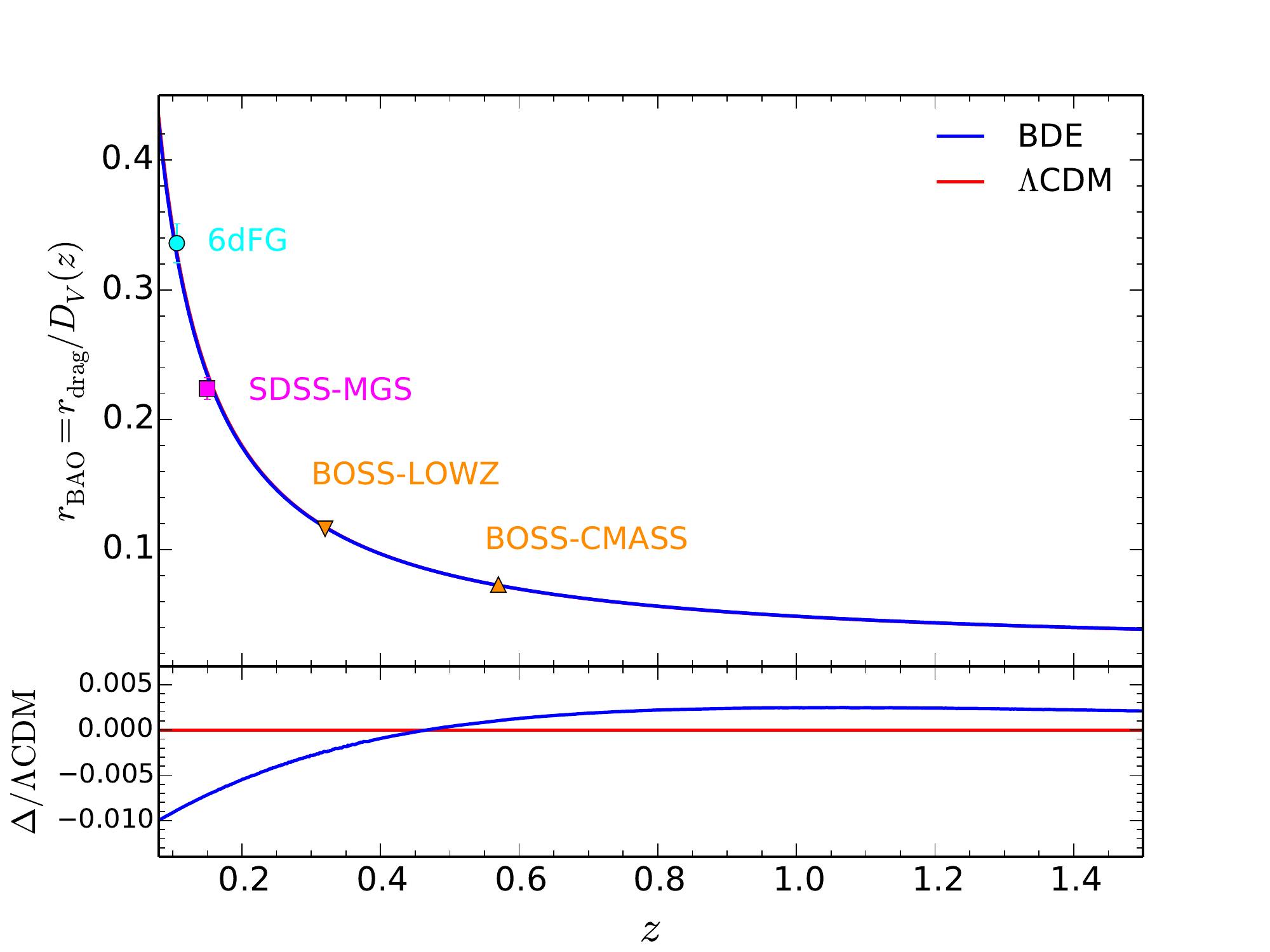}\caption{BAO ratio for the best fit BDE (blue) and $\Lambda$CDM (red) models. The markers are the observational points of the galaxy surveys we consider in this paper \cite{Beutler11_BAO,Ross15_BAO,GilMarin16_BAO}. The bottom panel shows the fractional difference $\Delta r_\textrm{BAO}/r_\textrm{BAO}$ w.r.t. $\Lambda$CDM.}\label{fig:implications_rbao}
\end{figure}

\begin{figure*}\centering                  
\includegraphics[width=1.\linewidth, height=0.22\textheight]{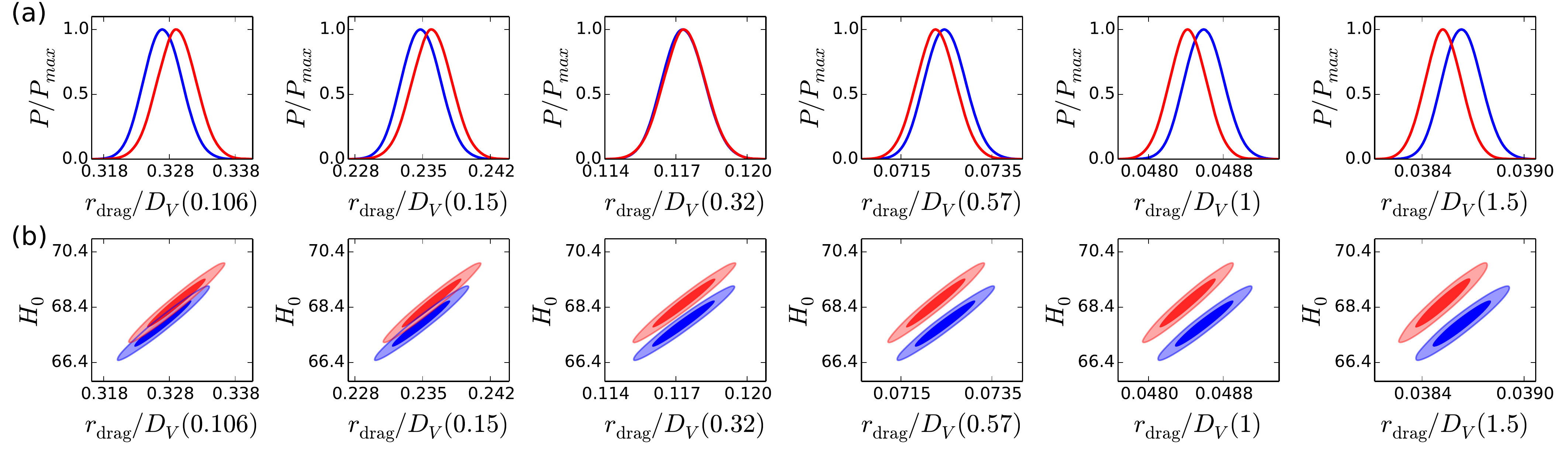}\caption{(a) Marginalised distributions in BDE (blue) and $\Lambda$CDM (red) of the BAO ratio at different redshifts. (b) Joint constraints on $H_0$ and the BAO ratio for each case. The contours cover the 68\% and 95\% confidence regions.}\label{fig:implications_bao}
\end{figure*}

The modification of the expansion rate at late times affects directly the cosmological distances probed by the SNeIa and BAO measurements as well as the position of the acoustic peaks in the CMB power spectrum. However, we remark that in order to fit the CMB data, there has to be a period of time where the expansion rate is smaller in BDE and a period of time where it is larger, as shown in the bottom panel of Fig. \ref{fig:implications_h}. This is because the CMB information tightly constrains \cite{PlanckCP13_CMB,PlanckCP15_CMB} the angular size of the sound horizon at recombination ($\theta_*$) ---which is one of the most precision measurements in cosmology, since its observational determination is less affected by the assumed cosmology and systematic effects \cite{PlanckCP13_CMB}--- which depends on the diameter distance $D_A(z_*)$ and the comoving sound horizon $r_*$ as $\theta_* = r_*/[(1+z_*)D_A(z_*)]$. Since $D_A$ is determined by the expansion history of the universe:
\begin{equation}\label{eq:implications_DA}
D_A=(1+z)^{-1}\int_{0}^{z}\frac{1}{H(z')}dz',
\end{equation}
and the constraints on $r_*$ are nearly the same, the differences in $H$ cancel each other, leading to a similar value of $D_A$ in both models as quoted in Table \ref{tab:mcmc_table}. That's why we get almost the same content of matter today in our model and more dark energy in $\Lambda$CDM. Therefore, this implies that in the very low$-z$ region where $H(\textrm{BDE})<H(\Lambda\mathrm{CDM})$, the differences in $H$ are not compensated in Eq. (\ref{eq:implications_DA}) and therefore here is where we can expect to observe the largest deviations w.r.t. $\Lambda$CDM. Figure \ref{fig:implications_rbao} shows the BAO ratio ($r_\textrm{BAO}$) for the best fit BDE and $\Lambda$CDM models:
\begin{equation} \label{eq:implications_rBAO_1}
 r_{\textrm{BAO}}(z)=\frac{r_{\textrm{drag}}}{D_V(z)}, 
\end{equation}
\begin{equation} \label{eq:implications_rBAO_2}
D_V(z) \equiv \left[ (1+z)^2 D_A^2(z)\frac{z}{H(z)} \right]^{1/3}, 
\end{equation}
where $r_\textrm{drag}$ is the comoving sound horizon at the drag epoch. Since $r_\textrm{drag}$ is very similar in both models (c.f. Table \ref{tab:mcmc_table}), the difference in $r_\textrm{BAO}$ depends basically on $H$ and its effects on $D_A$ through Eq. (\ref{eq:implications_DA}). As we mentioned before, $H(\textrm{BDE})<H(\Lambda\mathrm{CDM})$ for $z\leqslant0.24$, leading to $D_A(\textrm{BDE})>D_A(\Lambda\mathrm{CDM})$ and therefore to $r_\textrm{BAO}(\textrm{BDE})<r_\textrm{BAO}(\Lambda\mathrm{CDM})$ in this range. At $z\approx 0.32$ the expansion rate is nearly the same in both models $H(\textrm{BDE}) \approx H(\Lambda\mathrm{CDM})$, so $D_A$ is still larger in BDE and $r_\textrm{BAO}(\textrm{BDE})<r_\textrm{BAO}(\Lambda\mathrm{CDM})$, but the difference becomes small. Then, at higher redshifts where $H(\textrm{BDE})>H(\Lambda\mathrm{CDM})$, the diameter distance is now larger in $\Lambda$CDM giving $r_\textrm{BAO}(\textrm{BDE})>r_\textrm{BAO}(\Lambda\mathrm{CDM})$ as shown in the bottom panel of Fig. \ref{fig:implications_rbao}. We previously mentioned that our BDE model improves the fit to BAO measurements by $\chi^2_\textrm{BAO}(\textrm{BDE})/\chi^2_\textrm{BAO}(\Lambda\textrm{CDM})=0.788$, but now we can use these results in combination with the constraints on $H_0$ to find interesting deviations w.r.t. $\Lambda$CDM. Figure \ref{fig:implications_bao}(a) displays the marginalised distributions of $r_\textrm{BAO}$ at the effective redshifts of the BAO datasets we considered in our analysis as well as at $z=1$ and $z=1.5$. We see how the BDE distributions are shifted to the left w.r.t. $\Lambda$CDM at $z=0.106$ and $z=0.15$. Then, the distributions almost overlap at $z=0.32$ and finally the BDE contours are shifted to the right as expected. If we now draw the joint constraints with $H_0$ as in Fig. \ref{fig:implications_bao}(b), the contours split apart as the distributions of $r_\textrm{BAO}$ in BDE move to the right, leading to tensions w.r.t. $\Lambda$CDM at more than $2\sigma$. This tensions might be useful to test the dynamics of the dark energy and discriminate models in future galaxy surveys \cite{DESI,LSST,Euclid}.

\subsection{Evolution of matter perturbations.}\label{implications_deltam}
Matter perturbations are interestingly affected in our model since in this case the imprints left by the dark energy manifest not only at late times as expected, but also in the early universe when the condensation occurs. Here our model predicts a distinctive imprint on small-scale perturbations. 

The process of large scale structure formation is dominated by the dynamics of the cold dark matter \cite{Cooray02_SF,Amendola10_DE}. As we stated before, here we explore the linear regime where the fluctuations around the background are small enough to be studied with linear perturbation theory \cite{MaBertschinger95_Cosmo,HuSugiyama96_Cosmo,Amendola10_DE}. In that case, the evolution of CDM perturbations obey the equation of motion:
\begin{equation}\label{eq:implications_deltac}
 \delta_c''+\mathcal{H}\delta_c'-\frac{3}{2}\mathcal{H}^2\sum_i \Omega_i \delta_i (3c_{s,i}^2+1)=0,
\end{equation}
where the sum runs over all the fluids with sound speed $c_{s,i}^2=\delta P_i/\delta \rho_i$. We see that the dark energy has a twofold effect on matter perturbations: firstly at the background level through the modification of the expansion rate and secondly by the addition of an extra source term proportional to the dark energy inhomogeneities. We solve the complete set of perturbation equations \cite{Lewis00_CAMB} and for the sake of our analysis, we convert the total matter overdensities (baryons + CDM) into the newtonian gauge defined \cite{MaBertschinger95_Cosmo} by the gravitational potential $\Psi$ and the spatial curvature perturbation $\Phi$ by the line element $ds^2=a^2(\eta)[-(1+2\Psi)d\eta^2+(1-2\Phi)\delta_{ij}dx^idx^j]$.
\begin{figure}[b]\centering \includegraphics[width=1.0\linewidth]{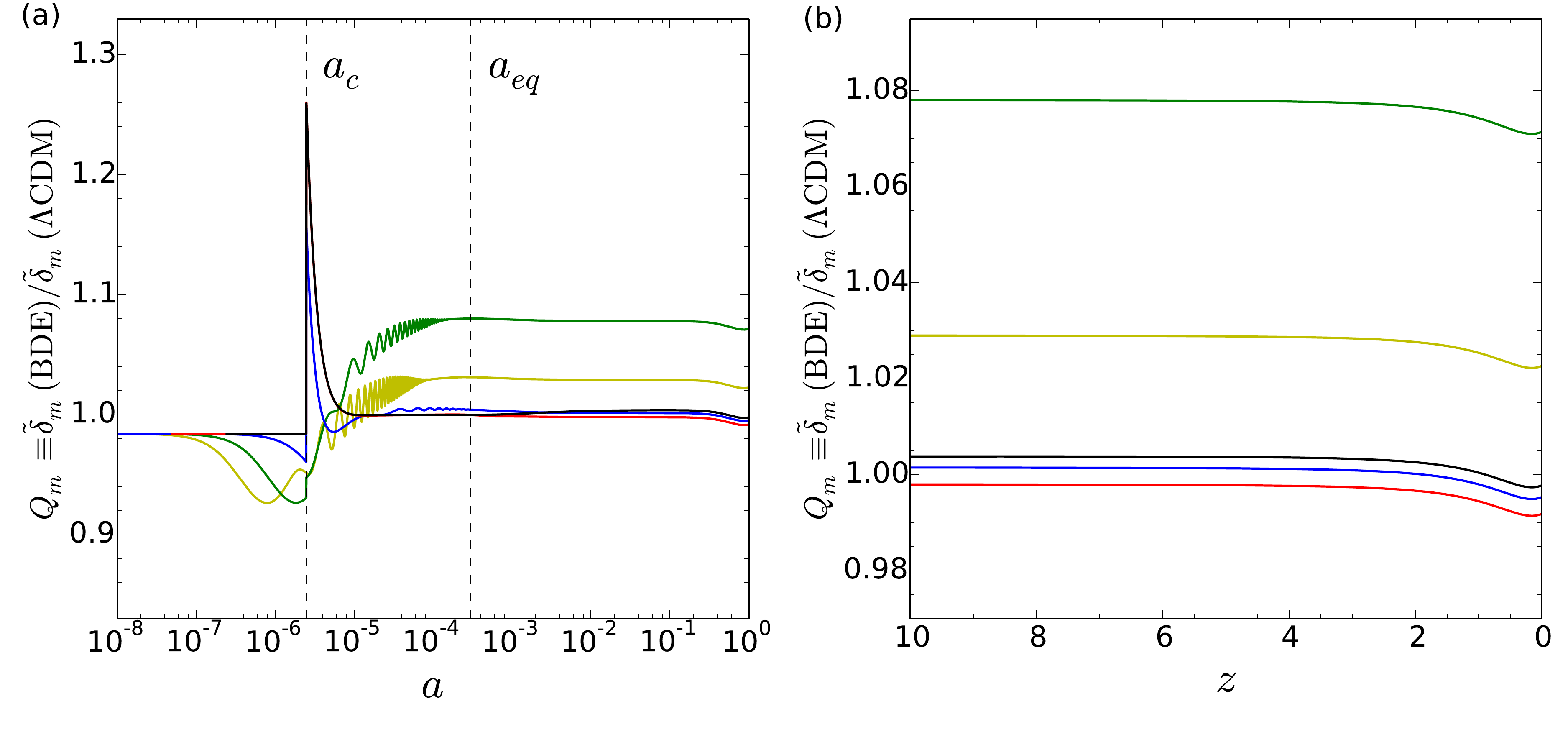}\caption{(a) Ratio of the matter overdensities in BDE w.r.t. $\Lambda$CDM for the best fit models for $k=10\mathrm{Mpc}^{-1}$ (yellow), $4.3\mathrm{Mpc}^{-1}$ (green), $1\mathrm{Mpc}^{-1}$ (blue), $0.05\mathrm{Mpc}^{-1}$ (red), and $0.005\mathrm{Mpc}^{-1}$ (black). $\delta_m$ has been converted into the newtonian gauge and normalised to the square root of the primordial power spectrum $P_s$ as  $\tilde{\delta}_m(k,a)=\delta_m(k,a)/\sqrt{P_s}$. The vertical dashed lines mark the condensation epoch $a_c$ and the matter-radiation equality era $a_{eq}$, respectively. (b) Ratio of the matter overdensities at late times for the modes shown in the left panel.}\label{fig:implications_deltam}
\end{figure}

Figure \ref{fig:implications_deltam}(a) shows the ratio of the matter overdensities ---normalised to the square root of the primordial power spectrum $P_s$--- for modes entering the horizon before, near and after the condensation for the best fit models:
\begin{equation}\label{eq:implications_deltam_normalised}
 \tilde{\delta}_m(k,a)=\frac{\delta_m(k,a)}{\sqrt{P_s}},
\end{equation}
where $P_s=A_s(k/k_0)^{n_s-1}$, with $k_0=0.05\textrm{Mpc}^{-1}$. Initially, when all modes are outside the horizon, matter perturbations do not evolve over time as expected in this gauge \cite{MaBertschinger95_Cosmo,HuSugiyama96_Cosmo}, leaving the ratio $Q_m\equiv \tilde{\delta}_m(\textrm{BDE})/\tilde{\delta}_m(\Lambda\textrm{CDM})$ constant as shown on the left side of the plot. However, matter perturbations in BDE are initially suppressed because of the DG affecting the initial amplitude of the matter perturbations through the gravitational potential $\Psi$, which depends on the fraction of relativistic particles besides the photons $R_\nu = \bar{\rho}_{eff}/(\bar{\rho}_{eff}+\bar{\rho}_{\gamma})$ \cite{MaBertschinger95_Cosmo}. Since $\bar{\rho}_{eff}^{\Lambda\textrm{CDM}}=\bar{\rho}_\nu$ and $\bar{\rho}_{eff}^\textrm{BDE}=\bar{\rho}_\nu + \bar{\rho}_\textrm{DG}$, the suppression factor $Q_{ini}=(1+(4/15)R_\nu^{\Lambda\textrm{CDM}})/(1+(4/15)R_\nu^{\textrm{BDE}})=0.984$ depends solely on $N_{ext}$ (c.f. Eq. (\ref{eq:bde_Next})) and it cannot be compensated by varying other cosmological parameters. 

Modes start evolving once they cross the horizon after $a_h$ defined implicitly in terms of the Hubble radius by $k=a_hH(a_h)$. However, there is a marked difference depending on whether or not the crossing epoch for a given mode $k$ is before the condensation, or equivalently, $k>k_c$, where $k_c=a_cH_\textrm{BDE}(a_c)=0.925\textrm{Mpc}^{-1}$ is the mode that enters the horizon just at $a_c$. Small modes $k>k_c$ are further suppressed w.r.t. $\Lambda$CDM  because the crossing time in this case is delayed by the presence of the DG. To see this, we solve $k=(a_hH(a_h))\big|_\textrm{BDE}=(a_hH(a_h))\big|_{\Lambda\textrm{CDM}}$ using Eqs. (\ref{eq:bde_rhoDGrhor_ac_2}) and (\ref{eq:bde_Friedmann_DG}), neglecting matter and $\rho_\Lambda$:
\begin{equation}\label{eq:implications_ah}
 \frac{a_h^\mathrm{BDE}}{a_h^{\Lambda\mathrm{CDM}}}=\sqrt{\frac{1-R_\nu^{\Lambda\mathrm{CDM}}}{1-R_\nu^\mathrm{BDE}}}=\sqrt{1+\frac{\rho_\mathrm{DG}(a_c)}{\rho_{r}(a_c)}}=1.062.
\end{equation}
Consequently, these modes cross the horizon earlier in $\Lambda$CDM, having more time to grow and therefore reducing further the ratio $Q_m$ as seen in the plot of Fig. \ref{fig:implications_deltam}(a). Nevertheless, this suppression effect is halted and subsequently reversed once the modes cross the horizon in BDE. Since at these times the universe is radiation dominated, matter perturbations evolve as $\delta_m \propto \ln a$ \cite{HuSugiyama96_Cosmo} and therefore the growth function $f\equiv \tfrac{d\ln\delta_m}{d\ln a}$ in this period is $f\propto 1/\delta_m$. Here $\tilde{\delta}_m(\textrm{BDE})< \tilde{\delta}_m(\Lambda\textrm{CDM})$, so $f(\textrm{BDE})> f(\Lambda\textrm{CDM})$ and hence $Q_m$ increases, leading to the troughs we see before $a_c$. 

\begin{figure}[t]\centering \includegraphics[width=1.\linewidth]{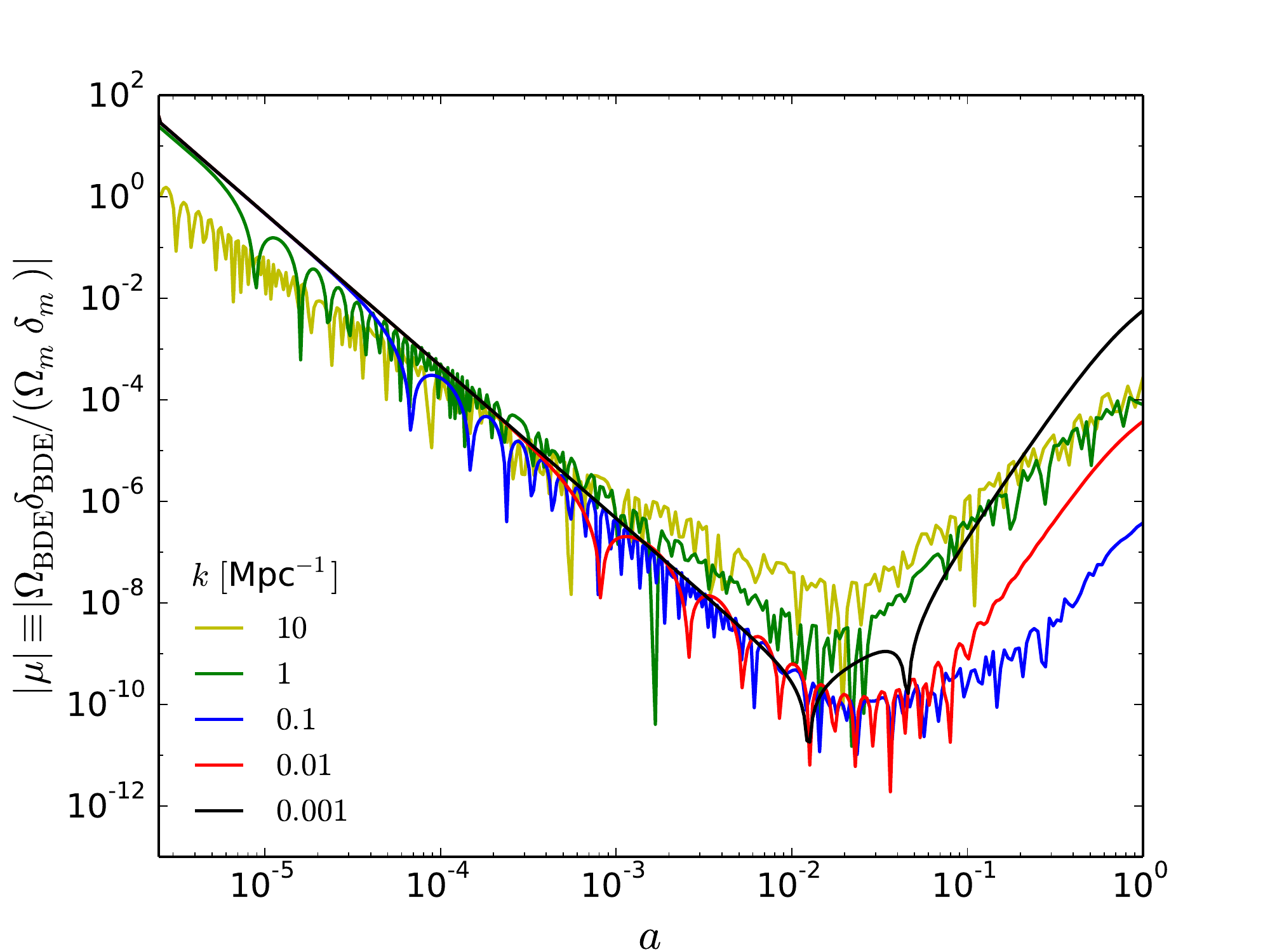}\caption{Ratio $\mu \equiv \Omega_\mathrm{BDE}\delta_\mathrm{BDE}/(\Omega_m \delta_m)$ of the dark energy to matter (baryons + CDM) perturbations in the newtonian gauge for the best fit BDE model. We plot the absolute value to show the dark energy fluctuations below the homogeneous background.}\label{fig:implications_mu}
\end{figure}

When the DG particles condense into the BDE meson, the expansion rate is now given by Eq. (\ref{eq:bde_Friedmann_phi}), where the dominant contribution at these times come from the standard radiation while the scalar field initially amounts to the 11\% of the energy content of the universe. However, we recall that the EoS leaps abruptly to $1$ and therefore the scalar field dilutes rapidly as $\rho_\textrm{BDE} \propto a^{-3(1+w_\textrm{BDE})}=a^{-6}$, leaving only the radiation as the dominant component. Although the expansion rate $H$ in BDE is still larger than in $\Lambda$CDM during the scalar field dilution, the deceleration $\ddot{a}/a=-8\pi G (2\rho_r+4\rho_\textrm{BDE})/3$ proceeds more efficiently since it includes the contribution from the scalar field. The extra deceleration in BDE increases the growth rate of matter perturbations compensating the initial suppression and boosting the ratio $Q_m$ above 1 as shown in the plot in Fig. \ref{fig:implications_deltam}(a). This effect is prominent since the energy density of BDE at $a_c$ is non-negligible and its dilution proceeds quickly. The final enhancement is mode-dependent reaching its maximum at $k\approx 4.3\textrm{Mpc}^{-1}$ and it completely stops once the scalar field has diluted enough before the matter-radiation equality epoch. This is the characteristic signature our model predicts on the linear evolution of matter perturbations for small modes $k>k_c$ crossing the horizon before the condensation. However, we remark that these modes also enter the non-linear regime earlier and it remains to see how much of this signature is affected by the non-linear dynamics \cite{Almaraz18_SF}. 

\begin{figure}[t]\centering \includegraphics[width=1.\linewidth]{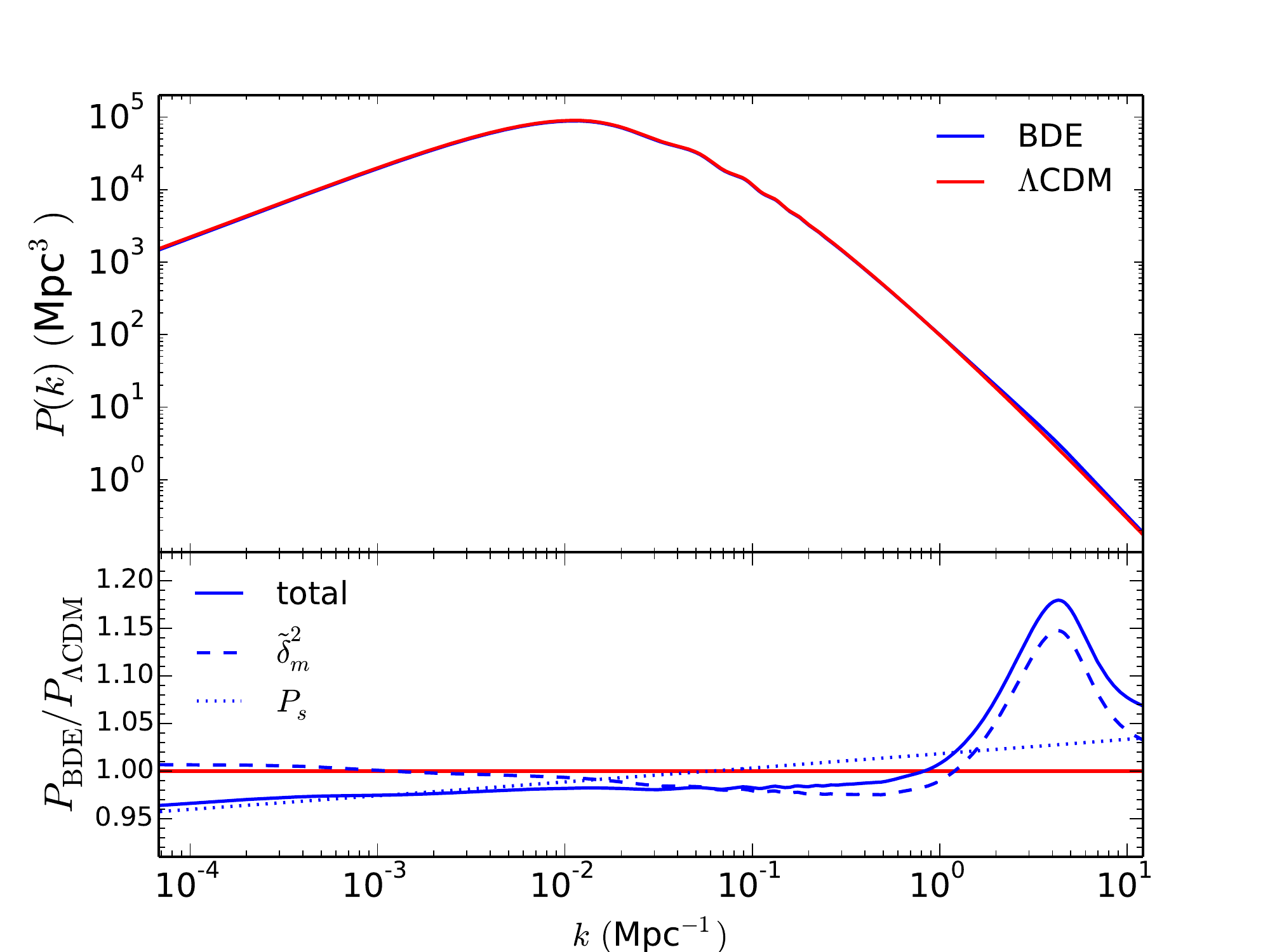}\caption{Matter power spectrum at $z=0$ for the best fit BDE (blue) and $\Lambda$CDM (red) models. The bottom panel shows the ratio w.r.t. $\Lambda$CDM of the total spectrum $P$ (solid) and the contributions from the primordial spectrum $P_s$ (dotted) and the matter overdensities $\tilde{\delta}_m$ (dashed).}\label{fig:implications_pk}
\end{figure}

\begin{figure*}\centering                  
\includegraphics[width=1.0\linewidth, height=0.22\textheight]{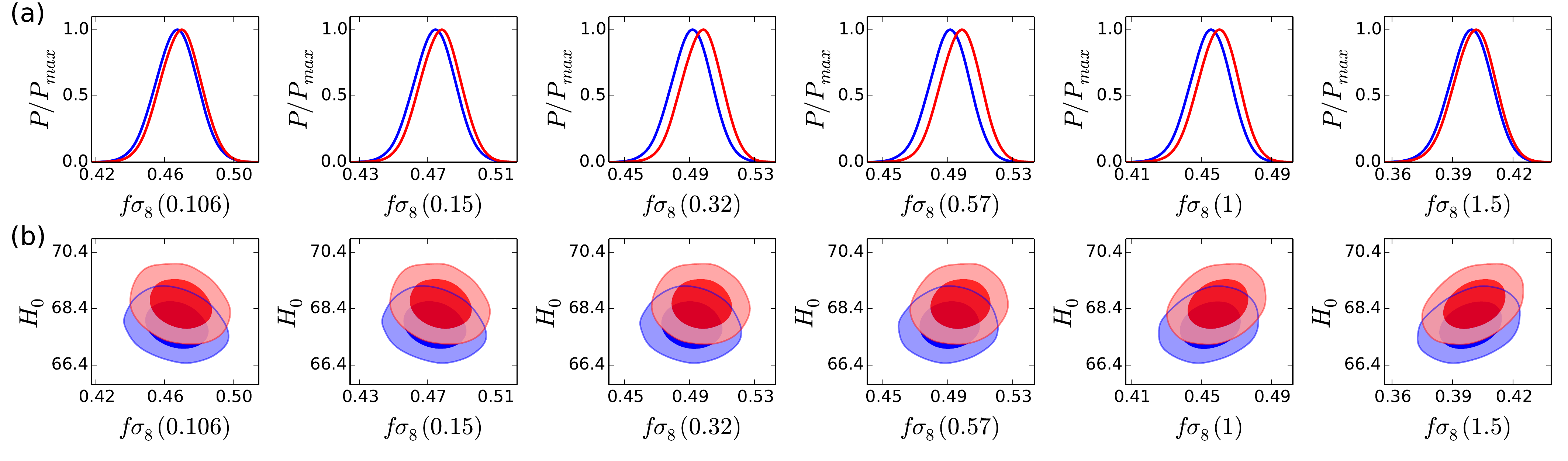}\caption{(a) Marginalised distributions in BDE (blue) and $\Lambda$CDM (red) of the combination $f\sigma_8$ at different redshifts. (b) Joint constraints on $H_0$ and $f\sigma_8$ for each case. The contours cover the 68\% and 95\% confidence regions.}\label{fig:implications_fsigma8}
\end{figure*}

Well within the matter domination era, the matter overdensities evolve as $\delta_m \propto a$ \cite{HuSugiyama96_Cosmo} in both models and consequently $Q_m$ is constant during this period. Finally, when the dark energy becomes dominant, the different late-time dynamics of the dark energy and its perturbations in BDE modify the growth rate of matter overdensities suppressing $Q_m$ a little bit as shown in detail in Fig. \ref{fig:implications_deltam}(b). We could have anticipated this result since between $0.24\leqslant z \leqslant 2.4$ the expansion rate is larger in BDE (c.f. Fig. \ref{fig:implications_h}), thus reducing the growth rate of matter perturbations in this period. Later, when the dark energy in BDE dilutes quickly for $z\leqslant 0.24$, the expansion rate is now larger in $\Lambda$CDM and the growth rate reduces in this case,making the curves end with a positive slope at $z=0$ as seen in the plot. The suppression factor is nearly the same for all the modes, dropping $Q_m$ by 0.62\% w.r.t. its constant value during matter domination with equal contributions from the dark energy background and its inhomogeneities. Actually, dark energy inhomogeneities affect the growth of matter perturbations inside the horizon by adding an extra source term in the Poisson equation \cite{Batista13_SF,Batista14_SF}:
\begin{eqnarray}\label{eq:implications_mu}
 k^2\Phi & \simeq & 4\pi G a^2(\bar{\rho}_m\delta_m+\bar{\rho}_\mathrm{BDE}\delta_\mathrm{BDE})\nonumber\\
& = &\frac{3}{2}\mathcal{H}^2\Omega_m \delta_m \left( 1+ \mu \right),
\end{eqnarray}
where $\mu \equiv \Omega_\mathrm{BDE}\delta_\mathrm{BDE}/(\Omega_m \delta_m)$. Figure \ref{fig:implications_mu} shows the evolution of this extra term $\mu$. Once the modes enter the horizon, dark energy perturbations fluctuate around the smooth background producing the wiggles we see in the plot. The size of the dark energy and the matter perturbations is comparable just after the condensation, but the ratio $\mu$ immediately decreases once the scalar field dilutes, reaching a minimum value around $a\sim 10^{-2}$. At late times the ratio $\mu$ grows, but still the modifications to the Poisson equation coming from the dark energy perturbations are very small, ranging between $\mathcal{O}(10^{-7}-10^{-5}$) today. On the other hand, modes crossing the horizon near and after the condensation are also affected by the rapid dilution of BDE, but the effect is transient and leaves the perturbations with almost the same amplitude as in $\Lambda$CDM before the matter-radiation equality epoch arrives. At late times these modes are suppressed by the dynamics of the dark energy and its perturbations too. 

The total power spectrum $P(k,a)=2\pi^2P_s|\tilde{\delta}_m(k,a)|^2/k^3$ is a combination of $\tilde{\delta}_m(k,a)$ affected by the dynamical processes described above, and the primordial spectrum $P_s$ determined by the values of the amplitude $A_s$ and tilt $n_s$ parameters. Figure \ref{fig:implications_pk} shows the power spectrum at $z=0$ for the best fit models. We can find out where the differences between BDE and $\Lambda$CDM arise by taking the ratio $P(\textrm{BDE})/P(\Lambda\textrm{CDM})$ as shown in the bottom panel. At large scales $k\lesssim 5\times 10^{-3}\mathrm{Mpc}^{-1}$ the deviations ($\leqslant 3.6\%$) w.r.t $\Lambda$CDM  mostly arise from the distinct primordial spectrum, which is not surprising since these modes cross the horizon lately having less time to evolve. On the other hand, the deviations at small scales $k\gtrsim 6\times 10^{-2}\mathrm{Mpc}^{-1}$ come from the dynamics. Here we can see the imprint left by the rapid dilution of BDE on modes $k>k_c$ entering the horizon before $a_c$ as the peak centred at $k\approx 4.3\mathrm{Mpc}^{-1}$, where the enhancement of power in BDE is about 18\%. However, we remark again that these modes cross the horizon at early times and today they are no longer in the linear regime. A non-linear approach must be used and we reserve for a future analysis how much of this discrepancy is expected to be seen when we take into account the non-linear effects \cite{Almaraz18_SF}.\\\indent 
\begin{figure}[b]\centering \includegraphics[width=1.\linewidth]{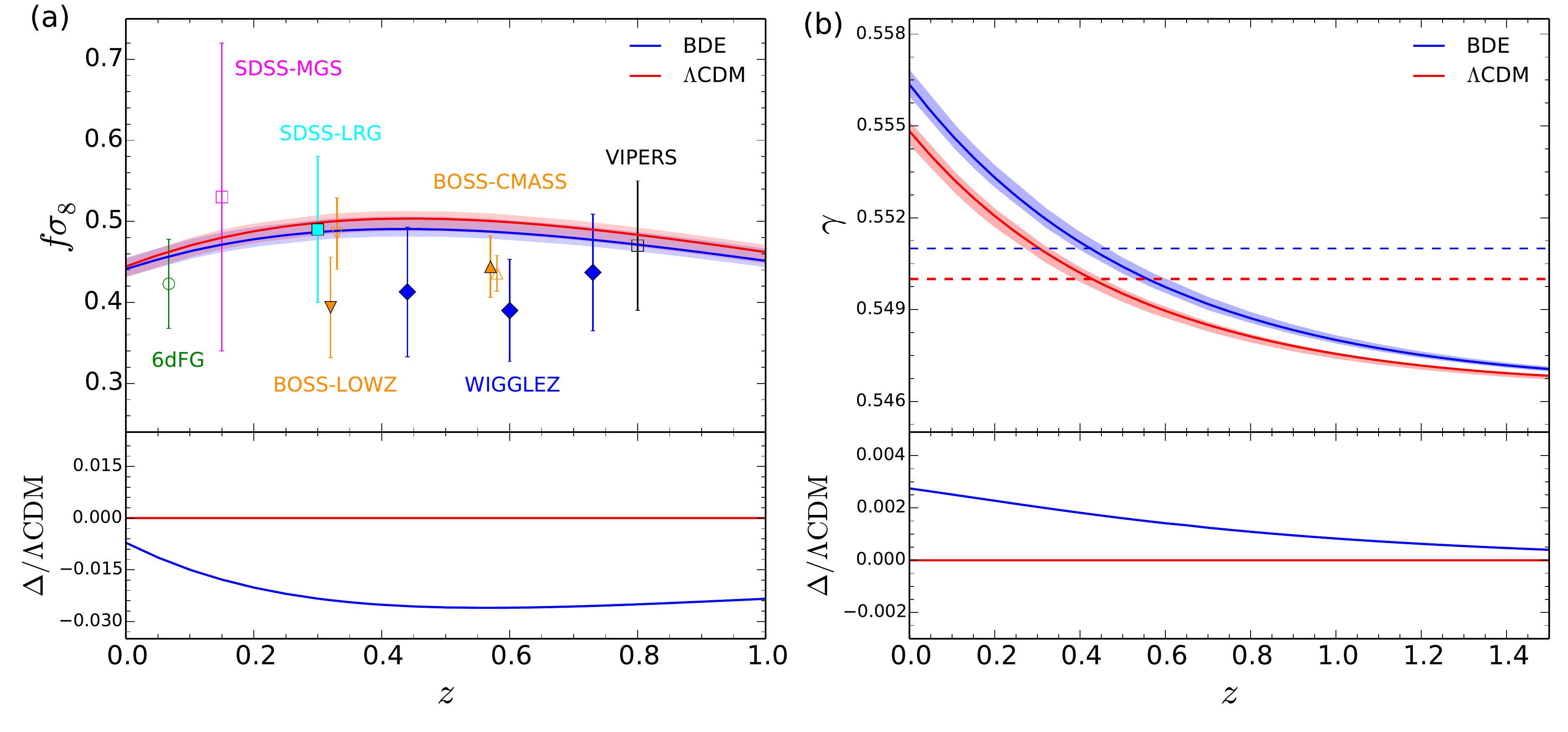}\caption{(a) Predictions on $f\sigma_8$ for BDE (blue) and $\Lambda$CDM (red). The solid curves correspond to the best fit models, while the bands show the 68\% confidence regios. We append the observational points measured by recent galaxy maps \cite{Beutler12_SF, Blake12_SF, DelaTorre13_SF, Oka14_SF, Howlett15_SF, GilMarin16_SF}. (b) Growth index ($\gamma$) for BDE (blue) and $\Lambda$CDM (red). The solid curves correspond to the best fit models, while in this case the bands show the 95\% confidence regions. The horizontal dashed lines are the $\gamma \simeq const$ approximation from the fitting formula $\gamma = 0.55 + 0.05[1+w(z=1)]$ \cite{Linder05_SF}. In both figures, the bottom panels show the relative difference of the best fit w.r.t. $\Lambda$CDM.}\label{fig:implications_fsigma8_gamma}
\end{figure}
Let's now consider consider the constraints on the parameters that characterise structure formation. The power spectrum is related to $\sigma_8$ ---the root mean square matter fluctuations in a sphere of radius $R=8h^{-1}\textrm{Mpc}$--- by $\sigma_8^2(z)= \frac{1}{2\pi^2}\int_0^\infty dlnk \textrm{ }k^3P(k,z) |W_8(k)|^2$, where $W_8(k)$ is the Fourier transform of the top-hat spherical function for this radius. The product $k^3P$ is suppressed below $k<10^{-3}\textrm{Mpc}^{-1}$ by the cubic power of $k$ and above $k>10^{-1}\textrm{Mpc}^{-1}$ by the window function, so the dominant contribution to $\sigma_8$ comes from the modes in the range $10^{-3} \textrm{ Mpc}^{-1} \leqslant k \leqslant 10^{-1} \textrm{ Mpc}^{-1}$, where $|W_8(k)|^2\simeq const$ and $P(\textrm{BDE})<P(\Lambda\textrm{CDM})$ according to Fig. \ref{fig:implications_pk}. We obtain $\sigma_8^{\textrm{BDE}}(z=0)=0.855$ and $\sigma_8^{\Lambda\textrm{CDM}}(z=0)=0.871$ for the best fit, which amounts to a deviation w.r.t $\Lambda$CDM of 1.8\%. As we move to higher redshifts, this difference is slightly reduced to 1.4\% at $z=1.5$. However, such a small deviation is still beyond the accuracy of the current measurements \cite{Abate09_SF,Ettori10_SF,*Liu15_SF}. Moreover, if we consider the whole set of MCMC samples the marginal limits agree at the $1\sigma$ level in this range, particularly at $z=0$, where we obtain $\sigma_8^{\textrm{BDE}}(z=0)=0.861 \pm 0.022$ and $\sigma_8^{\Lambda\textrm{CDM}}(z=0)=0.864\pm 0.022$.

Measurements of redshift space distortions (RSD) constrain the combination $f\sigma_8$. Figure \ref{fig:implications_fsigma8} shows the marginal distributions and the joint constraints with $H_0$ for the same redshifts as in Fig. \ref{fig:implications_bao}. Even though the confidence limits agree within the $1\sigma$ level, we see that $f\sigma_8(\textrm{BDE})<f\sigma_8(\Lambda\textrm{CDM})$. However, the difference is very small and there is no a marked correlation with $H_0$ as occurs in the BAO ratio (c.f. Fig. \ref{fig:implications_bao}(b)). Therefore, we don't observe any tension w.r.t. $\Lambda$CDM in this case. The evolution of $f\sigma_8$ at late times is displayed in Fig. \ref{fig:implications_fsigma8_gamma}(a), where we include the measurements of recent galaxy maps \cite{Beutler12_SF, Blake12_SF, DelaTorre13_SF, Oka14_SF, Howlett15_SF, GilMarin16_SF}. For the best fit, the difference is about 2.6\% in $0.4\leqslant z \leqslant 0.8$. Current research aimed to reduce the error bars of the observational points studies the impact of nonlinearities and the subtraction of the Alcock-Paczynski effect \cite{GilMarin16_SF,Li18_SF}. Galaxy maps also provide information on the growth index $\gamma$ defined by \cite{Peebles80_Cosmo}: $f=\Omega_m^\gamma(a)$. Figure \ref{fig:implications_fsigma8_gamma}(b) shows the best fit curves and the 95\% confidence regions obtained from our MCMC analysis. Here we find a marked discrepancy $\gamma(\textrm{BDE})>\gamma(\Lambda\textrm{CDM})$ especially at late times, where the difference w.r.t. $\Lambda$CDM amounts to 0.3\% at $z=0$ for the best fit. We see that although $\gamma$ varies over the time, its time dependence is very mild and so $\gamma\simeq const$ is a good approximation as occurs in smooth dark energy models \cite{Linder05_SF,Batista14_SF}. In this respect, the horizontal dashed lines correspond to $\gamma(\Lambda\textrm{CDM})=0.550$ and $\gamma(\textrm{BDE})=0.550\pm 0.001$ obtained from the fitting formula \cite{Linder05_SF}: $\gamma = 0.55 + 0.05[1+w(z=1)]$, where we use $w_\textrm{BDE}(z=1)=-0.98$ according to Fig. \ref{fig:mcmc_bde_background}(b). The difference is now 0.18\% w.r.t. $\Lambda$CDM. In any case, these deviations are beyond the precision limits of the observational data at present time \cite{Abate09_SF,Beutler12_SF,Howlett15_SF}. 

\subsection{Light element abundances.}
The extra amount of dark radiation due to DG enhances the expansion rate of the universe before the condensation, particularly, during the Big Bang Nucleosynthesis (BBN) era. The larger expansion rate leads to a higher freeze-out temperature $T_f\simeq (H/G_F)^{1/5}$ (where $G_F$ is the Fermi constant) at which the neutrinos decouple from the electrons and positrons, which in turn increases the neutron-to-proton ratio $(n/p)_f \simeq e^{-(m_n-m_p)/T_f}$ (where $m_n$ and $m_p$ is the neutron and the proton mass, respectively) just before the onset of BBN \cite{Cyburt16_BBN}. Since the amount of helium-4 produced during BBN is proportional to $(n/p)$ \cite{Cyburt16_BBN}, we expect an enhanced production of primordial helium in our  model. The last two parameters of Table \ref{tab:mcmc_table} show the abundances of primordial helium and deuterium that we found in our MCMC analysis. Although BDE predicts an excess of helium w.r.t $\Lambda$CDM of 4.9\% and an excess of 12\% of deuterium, our results are consistent with the abundances obtained from measurements of emission lines in HII regions \cite{Aver12_BBN,Aver13_BBN,*Izotov14_BBN} and quasar absorption systems \cite{Iocco09_BBN,*Cooke14_BBN,*Riemer15_BBN,*Zavarygin18_BBN}. However, here we remark that despite the progress made in the recent years, BBN abundances are still prone to systematic effects (such as the neutron lifetime \cite{Yue13_HEP,*Olive14_HEP}, the quality of the data \cite{Aver12_BBN}, among others) impeding the accurate determination both at theoretical and observational level \cite{Cyburt16_BBN}. 

\subsection{The CMB spectrum.}
\begin{figure}[t]\centering \includegraphics[width=1.\linewidth]{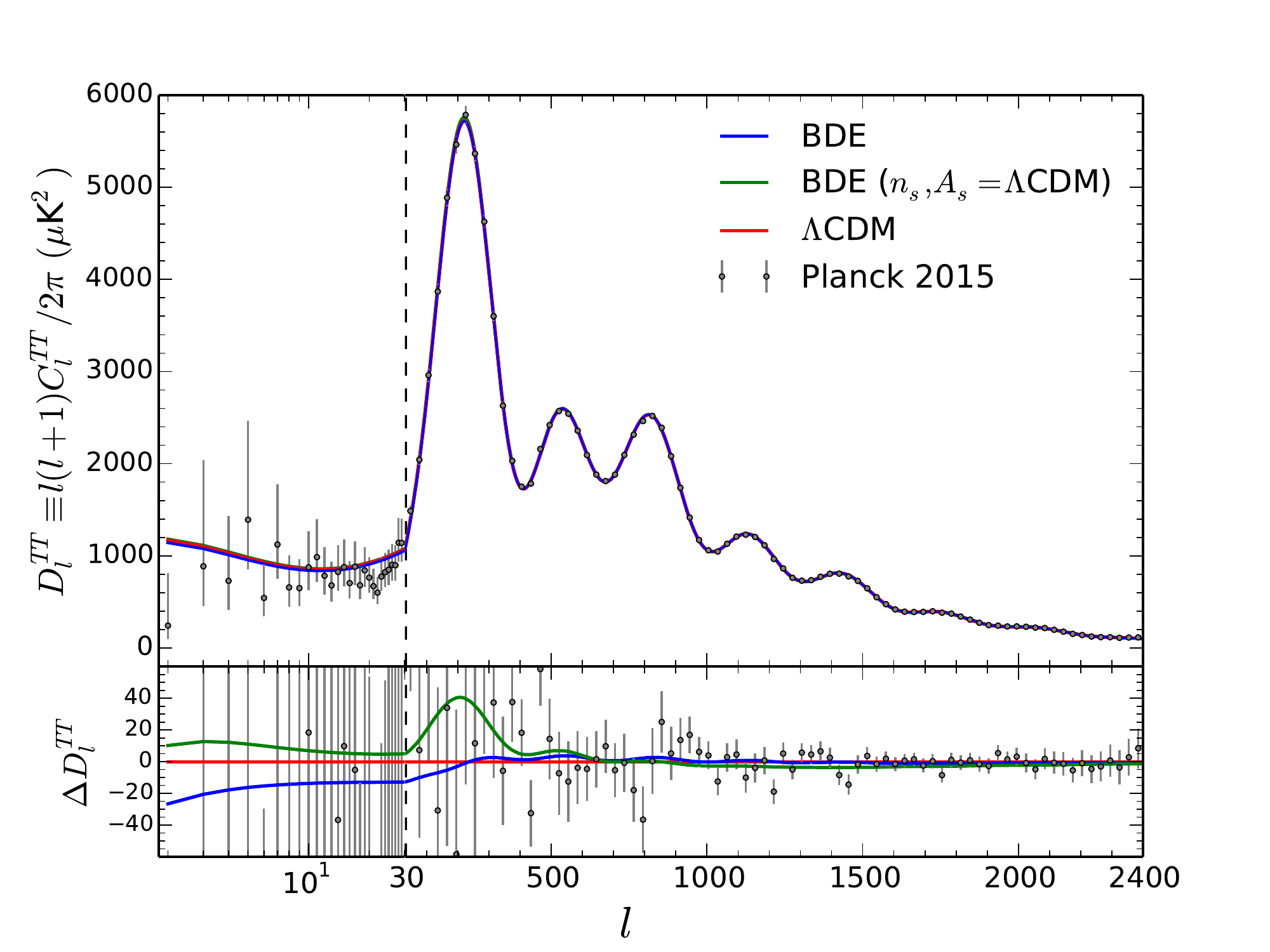}\caption{CMB temperature anisotropy spectrum for the best fit BDE (blue) and $\Lambda$CDM (red) models. The grey circles are the Planck 2015 measurements \cite{PlanckCP15_CMB}. The green curve is the spectrum for a BDE model where $n_s$ and $A_s$ are set equal to the best fit $\Lambda$CDM values while the other basic parameters are kept to the BDE best fit. The bottom panel shows the absolute residuals $\Delta D_l^{TT}\equiv D_l^{TT}(\mathrm{BDE})- D_l^{TT}(\Lambda\mathrm{CDM})$. We display the multipole scale using a logarithmic scale between $2\leqslant l \leqslant 30$ and then a linear scale from the vertical dashed line onwards.}\label{fig:implications_cmb}
\end{figure}

Our BDE model impacts the CMB temperature anisotropy spectrum in multiple ways and it is interesting to see how these effects combine together to fit the data. Figure \ref{fig:implications_cmb} shows the computed CMB spectra for the best fit BDE (blue) and $\Lambda$CDM (red) models with measurements from the Planck (2015) satellite \cite{PlanckCP15_CMB}. The bottom panel displays the residuals w.r.t. $\Lambda$CDM. The overall fit is equivalent for both models $\chi^2_\textrm{CMB}(\textrm{BDE})/\chi^2_\textrm{CMB}(\Lambda\textrm{CDM})\simeq 1$ as seen from Table \ref{tab:mcmc_table}. However, we find a slight preference $\chi^2_{lowl}(\textrm{BDE})/\chi^2_{lowl}(\Lambda\textrm{CDM})= 0.914$ for BDE in the low multipole region $l=2-29$, where the accuracy of the measurements is limited by the cosmic variance \cite{PlanckCP13_CMB,PlanckCP15_CMB}; for higher multipoles $30 \leqslant l \leqslant 2508$, $\chi^2_{highl}(\textrm{BDE})/\chi^2_{highl}$ $(\Lambda\textrm{CDM})= 1.001$. In general, dark energy affects the CMB power spectrum in the location of the acoustic peaks given by \cite{HuSugiyama95_CMB,Durrer08_CMB,Amendola10_DE}: $l_n=\tfrac{n\pi}{2r_*}(1+z_*)D_A(z_*)$, with $n=1,2,...$, and the late-time integrated Sachs-Wolfe (ISW) effect \cite{SachsWolfe67_Cosmo}, which depends on the evolving rate of the newtonian potentials. In a matter-dominated universe these potentials don't change in time and therefore the late-time ISW effect is absent, suppressing the CMB spectrum at low $l$ \cite{Amendola10_DE,HuSugiyama95_CMB}. 

When we consider the non-trivial dynamics of the dark energy in quintessence models, the background evolution and the dark energy inhomogeneities modify the ISW effect in opposite ways \cite{Weller03_QCDM,*Bean04_QCDM}: for quintessence fields whose EoS is always larger than $w_\Lambda=-1$ and therefore it doesn't cross the phantom regime ($w<-1$) there is a suppression of the ISW effect, while dark energy perturbations with a sound speed $c_s^2>0$ enhances the power. The dominant effect comes from the dark energy perturbations \cite{Weller03_QCDM,*Bean04_QCDM} and consequently these theories are characterised by more power at low $l$ as compared with a $\Lambda$CDM cosmology running with the same set of basic parameters. In our BDE model there is and additional imprint on the CMB left by the DG at high multipoles $l\gtrsim 1000$. The tail of the spectrum is damped by the factor $D_\gamma(k)=\int_0^{\eta_0}d\eta \textrm{ } \tau'e^{-\tau}e^{-[k/k_D(\eta)]^2}$, where $\tau$ is the optical depth and the wavenumber $k_D$ is directly proportional to the square root of the number density of free electrons $n_e$, $k_D\propto \sqrt{n_e}$ \cite{HuSugiyama95_CMB,Hu95_CMB,HuWhite97_CMB,Hou13_CMB}. Since the ionisation energy of helium is higher than hydrogen, the larger amount of helium produced in our BDE model traps more free electrons reducing $n_e$ before recombination starts and thus making $e^{-(k/k_D)^2}$ smaller, which in turn leads to a more damped tail of the spectrum \cite{Trotta04_CMB,*Ichikawa06_CMB}. We stress that the influence of the DG on the CMB is indirectly given by its effect on the helium abundance and not due to its presence at recombination. Since in our model the extra amount of dark radiation $N_{ext}=0.945$ vanishes at $a_c$ long before recombination, the DG is not present in any physical process taking place onwards, which is a completely different scenario from the usual extensions to $\Lambda$CDM where the extra amount of radiation $N_{eff}>3.046$ is held constant throughout the history of the universe. Therefore, the constraints on $N_{eff}$ found in these analyses \cite{Calabrese11_CMB,Hou13_CMB,PlanckCP13_CMB,PlanckCP15_CMB} don't apply for our model. 

When we fit the CMB data, the constraints on $z_*$, $r_*$ and $D_A(z_*)$ are very similar in both models, so there is no perceptible shift in the location of the acoustic peaks in Fig. \ref{fig:implications_cmb}. The damping wavenumber is smaller in BDE as expected, but $\tau(\textrm{BDE})<\tau(\Lambda\textrm{CDM})$ for the best fit spectra of the plot, which compensates the effect of $k_D$ on $D_\gamma$ and leads to no extra damping of the BDE spectrum tail. The only observable deviations w.r.t $\Lambda$CDM occur in the low$-l$ region, where BDE unexpectedly has less power. However, this effect is a consequence of the different $n_s$ and $A_s$ in each model. This can be seen in the green curve corresponding to a BDE model where $n_s$ and $A_s$ are the same as $\Lambda$CDM but the other basic parameters are kept to their best BDE best fit values. In this case we recover the enhanced power at low$-l$ produced by the dark energy perturbations. The small bump around the first peak at $l\approx 220$ is due to the larger amplitude of the early ISW effect produced by the delay of the matter-radiation equality epoch in BDE, $z_{eq}(\textrm{BDE})<z_{eq}(\Lambda\textrm{CDM})$ \cite{Cabass15_CMB}. For $l\gtrsim 900$ the spectrum is mildly suppressed. We see then that the effect of $n_s$ and $A_s$ on the fitted spectrum extends up to the first peak. The values of these parameters have to be properly adjusted to fit the height of the first peak. In doing so, the effect of the dark energy perturbations is washed out leaving the BDE spectrum below $\Lambda$CDM in this region.   

\section{Conclusions}\label{conclusions}
In this paper we present in detail the cosmological implications and observational constraints on the dynamical Bound Dark Energy model as an alternative to the cosmological constant $\Lambda$CDM model to explain the late-time acceleration of the universe. Our BDE model aims to explain the nature of the dark energy at fundamental level using a natural extension of the Standard Model of particle physics. We introduce a hidden dark gauge group of weakly coupled particles which undergo a phase transition at late times, forming a light scalar meson particle that represents the dark energy in our model. The evolution of this dark energy meson is described by a canonical scalar field $\phi$ with an IPL potential $V(\phi)=\Lambda^{4+2/3}\phi^{-2/3}$ with an exponent $n=2[1+2/(N_c-N_f)]=2/3$ which is obtained from a supersymmetric gauge group $SU(N_c=3)$, $N_f=6$ using the Affleck-Dine-Seiberg techniques \cite{ADS85_HEP}.

The Dark Energy corresponds to the lightest meson field $\phi$ formed once the gauge coupling constant of DG becomes strong, which takes place at the condensation energy scale $\Lambda_c$ and at a scale factor  $a_c$. The scalar potential $V$, the exponent $n=2/3$ and the condensation energy scale $\Lambda_c$ are all derived quantities depending only on the choice of the Dark Group. It is worth noticing that our DG has the same fundamental status as the well established Standard Model of particle physics---$SU_\textrm{QCD}(N_c=3)\times SU(N_c=2)_L\times U_Y(N_c=1)$ with 3 families describing the strong and electroweak interactions---in the sense that $N_c$ and $N_f$ are quantities not derived from a deeper theory. The BDE model has then no free parameters and the initial conditions of the scalar field naturally arise from physical considerations. Moreover, the evolution of the scalar field is completely determined by the dynamical equations and our BDE dark energy model has then one less free parameter than $\Lambda$CDM. However, the value of $\Lambda_c$ depends on the values of $\Lambda_{gut}$ and $g_{gut}$ which are still not well determined \cite{Bourilkov15_HEP}.

We constrain our model using recent measurements of the CMB temperature anisotropy spectrum, the luminosity distance of SNeIa and the BAO signal in galaxy maps at different redshifts. According to our results, the condensation of the BDE meson occurs at $a_c= (2.48\pm 0.02) \times 10^{-6}$ well within the radiation domination epoch at an energy scale of $\Lambda_c=44.09\pm 0.28 \textrm{ eV}$, which is not only consistent, but also improves our theoretical bound $\Lambda_c^{th}=34^{+16}_{-11} \textrm{ eV}$ based on high-energy physics information. The relation between $\Lambda_c$ and $a_c$ is remarkably verified by a small deviation of only $0.2\%$ w.r.t. the theoretical prediction $a_c\Lambda_c/\mathrm{eV}=1.0939\times 10^{-4}$ when we allow these two parameters to vary indenpendently. Moreover, the evolution of the EoS of the dark energy and the expansion rate at late times are insensitive to the initial value of the EoS at $a_c$. The BDE model is in excellent agreement with the observations, particularly with the BAO data, which in the near future are expected to provide key evidence on the dynamics of the dark energy. In this case, BDE fits better the observations by improving the likelihood by a ratio of 2.1 compared to $\Lambda\mathrm{CDM}$, reducing $(\chi^{2})^{\Lambda \mathrm{CDM}}_\mathrm{BAO}=7.115$ in $\Lambda$CDM to $(\chi^{2})^\mathrm{BDE}_\mathrm{BAO}=5.609$ in our BDE model.

The dynamics of the dark energy in our model leaves imprints on cosmological quantities within the reach of current observations. At late times the dilution of the dark energy driven by the EoS modifies the expansion rate and thus the cosmological distances probed by BAO measurements in galaxy maps and SNeIa surveys. In order to fit the location of the acoustic peaks in the CMB spectrum, BDE is required to have roughly the same amount of matter as in $\Lambda$CDM (a tiny excess of $0.5\%$ in $\Lambda$CDM), but less dark energy today by $3.7\%$. This difference can be used in combination with the BAO constraints to find tensions w.r.t. $\Lambda$CDM in the $r_\mathrm{BAO}-H_0$ plane, which may be explored by future missions such as DESI \cite{DESI}, LSST \cite{LSST} and Euclid \cite{Euclid}.

The growth rate of matter overdensities is interestingly affected at small scales. The amplitude of the modes that cross the horizon before $a_c$ is initially suppressed w.r.t. $\Lambda$CDM  by the free streaming of the particles of the DG. Then these modes are enhanced because of the rapid dilution of BDE just after the condensation, and finally there is a slight suppression because of the different dynamics and the contribution of the dark energy perturbations in our model. The resulting effect is an enhancement of the matter power spectrum in BDE at small scales, where the differences w.r.t. $\Lambda$CDM rise up to $18\%$ at $k\approx 4.3 \mathrm{ Mpc}^{-1}$. However, these modes are no longer in the linear regime and we reserve for a future work the analysis of how much of this signature is affected by the non-linear dynamics of structure formation \cite{Almaraz18_SF}.\\\indent
The presence of the DG prior the condensation of the BDE meson introduces an additional amount of radiation ($N_{ext}=0.945$ for $a_{\nu dec}<a<a_c$) in the early universe. However, this extra dark radiation vanishes at the phase transition ($N_{ext}=0$ for $a\geqslant a_c$) long before the decoupling era and therefore it has no direct influence on the CMB. Our BDE model affects the spectrum of temperature anisotropies in multiple ways, but when we fit the data, the largest deviations w.r.t. $\Lambda$CDM occurs in the low multipole region $l\leqslant 30$ sensitive to the cosmic variance. The DG increases the expansion rate in the early universe and therefore leads to an enhanced amount of primordial helium and deuterium produced at BBN. Here we find a marked difference w.r.t. $\Lambda$CDM, but our BDE model is also consistent with the astrophysical bounds and more precise measurements are required to draw any further conclusions.\\\indent
The problem of the dark energy is a challenging question whose solution requires the introduction of new physics. Any alternative scenario intended to replace the cosmological constant as the cause of the cosmic acceleration not only has to fit consistently the observations, but it must also lie on a solid theoretical foundation which allows us to understand what is the source of the dark energy and why it has come to drive the expansion of the universe at late times. Additionally, this candidate must predict deviations from the standard $\Lambda$CDM scenario that can be probed by future observations. Any reduction of free parameters and a better fit to cosmological observations compared to $\Lambda$CDM would give a solid hint on the nature of the dark energy. The dark energy model we present here fulfills all these requirements; BDE has a sound derivation as a natural extension of the Standard Model of particle physics, it has no free parameters and it has an excellent fit with current cosmological data, particularly improving the likelihood of the BAO distance measurements (specially designed to provide key evidence on the properties of the dark energy)  w.r.t. $\Lambda$CDM. With the advent of the precision era in cosmology and the quest of a new paradigm for the dark sector of the universe, our BDE model provides an interesting framework to be explored in the forthcoming years.

\begin{acknowledgments}
We acknowledge the financial support from projects UNAM PAPIIT IN103518 and Conacyt Fronteras 281.
\end{acknowledgments}

\appendix
\section{Fitting formula for the equation of state}\label{App}
\begin{figure}[t]\centering \includegraphics[width=1.\linewidth]{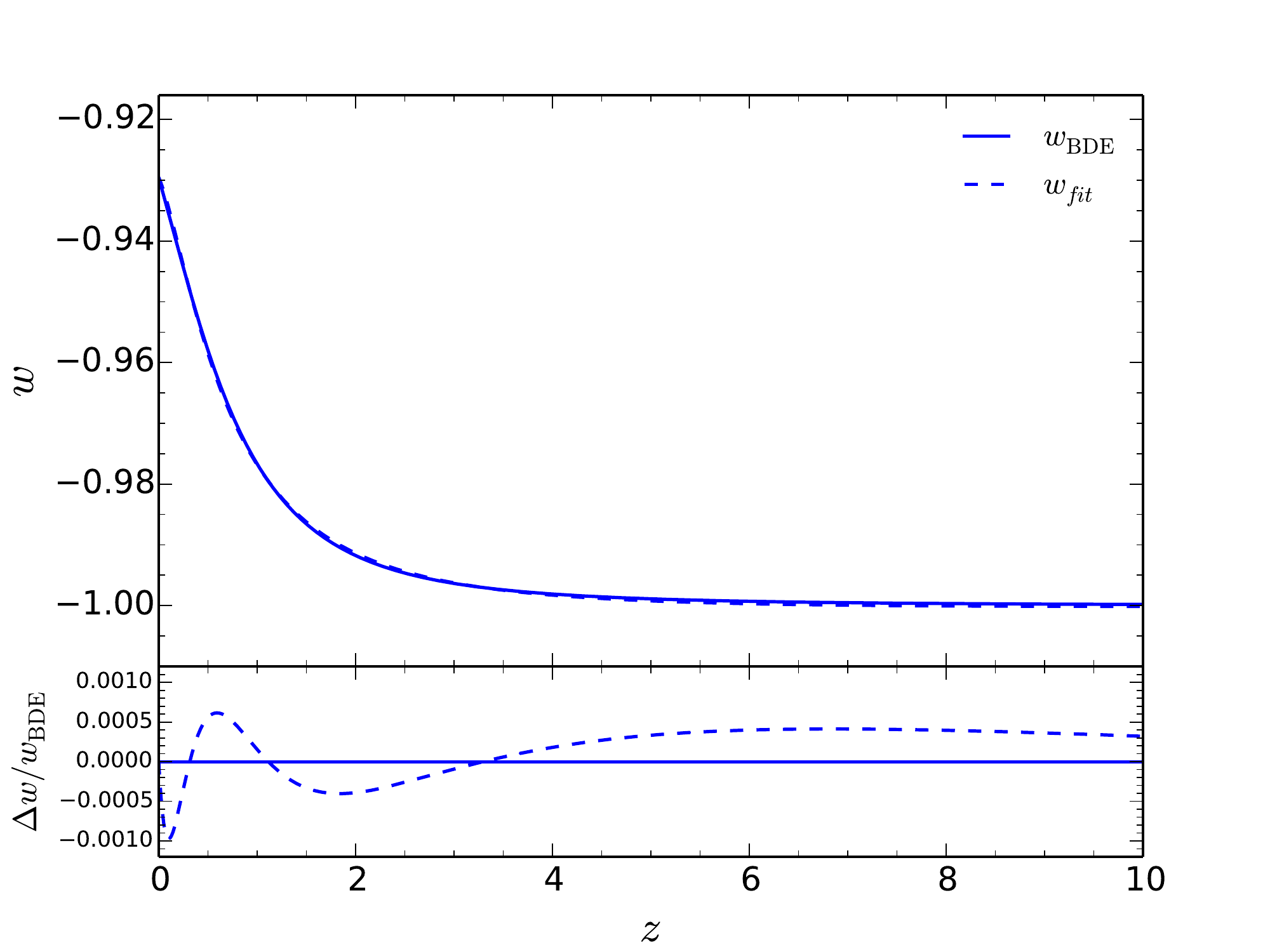}\caption{Evolution of the EoS at late times for the best fit BDE model (solid) and the fitting formula of Eq. (\ref{eq:appendix_wfit}) (dashed). The bottom panel shows the relative difference $\Delta w= (w_\textrm{fit}-w_\textrm{BDE})/w_\textrm{BDE}$.}\label{fig:appendix_eosfit}
\end{figure}
Instead of solving the Klein-Gordon (\ref{eq:bde_KG}) and the Friedmann (\ref{eq:bde_Friedmann_phi}) equations for the background, we can account for the evolution of the EoS at late times by using the fitting formula:
\begin{equation}\label{eq:appendix_wfit}
w_\textrm{fit}= \sum_{i=0}^4 \frac{b_iz^i}{(1+z)^4},
\end{equation}
where $w_\textrm{fit}(z=0)=b_0$ and $w_\textrm{fit}(z\gg 1)=b_4$, with fitting coefficients $b_0=-0.9296$, $b_1=-3.752$, $b_2=-5.926$, $b_3=-4.022$ and $b_4=-0.999$. 
Figure \ref{fig:appendix_eosfit} shows $w_{fit}$ and the best fit curve ($w_\mathrm{BDE}$) obtained from our MCMC analysis. We see that the formula (\ref{eq:appendix_wfit}) provides an excellent fit, giving a relative error w.r.t. $w_\mathrm{BDE}$ below 0.1\% for $z<140$, shortly after the EoS drops from $\simeq 1$ in Fig. \ref{fig:mcmc_bde_background}(a). Therefore, this practical parametrization captures the late-time dynamics of the dark energy in our model, leaving unaltered the cosmological distances and giving the same suppression factor of matter overdensities, since the relevance of the dark energy perturbations at late times is small ($\mu \sim \mathcal{O}(10^{-7}-10^{-5})$), as we discussed in section (\ref{implications_deltam}).

\bibliography{bde_revtex}

\end{document}